\documentclass[11pt]{amsart}

\usepackage{amssymb}
\usepackage{epsfig}
\usepackage{comment}
\usepackage{amsmath}
\usepackage{subfigure}
\usepackage[section]{placeins}
\usepackage{mathrsfs}
\usepackage{graphicx}
\usepackage[notrig]{physics}
\usepackage{units}
\usepackage{pstricks}
\usepackage{color}
\usepackage{algorithm}
\usepackage{algorithmic}

\theoremstyle{definition}

\theoremstyle{remark}

\graphicspath{{./Figures/}}

\begin{document}

\title[The finite-particle method]{An optimal-transport finite-particle method for mass diffusion}

\author{
A.~Pandolfi${}^1$, L.~Stainier${}^2$ and M.~Ortiz${}^3$
}

\address
{
${}^1$Politecnico di Milano,
Civil and Environmental Engineering Department, 20133 Milano, Italy
\\
${}^2$Institut de Recherche en G{\'e}nie Civil et M{\'e}canique (GeM), Ecole Centrale de Nantes, 4321 Nantes cedex 3, France
\\
${}^3$California Institute of Technology,
Engineering and Applied Science Division, Pasadena CA, 91125, USA
}

\email{anna.pandolfi@polimi.it, laurent.stainier@ec-nantes.fr, ortiz@caltech.edu}

\begin{abstract}
We formulate a class of velocity-free finite-particle methods for mass transport problems based on a time-discrete incremental variational principle that combines entropy and the cost of particle transport, as measured by the Wasserstein metric. The incremental functional is further spatially discretized into finite particles, i.~e., particles characterized by a fixed spatial profile of finite width, each carrying a fixed amount of mass. The motion of the particles is then governed by a competition between the cost of transport, that aims to keep the particles fixed, and entropy maximization, that aims to spread the particles so as to increase the entropy of the system. We show how the optimal width of the particles can be determined variationally by minimization of the governing incremental functional. Using this variational principle, we derive optimal scaling relations between the width of the particles, their number and the size of the domain. We also address matters of implementation including the acceleration of the computation of diffusive forces by exploiting the Gaussian decay of the particle profiles and by instituting fast nearest-neighbor searches. We demonstrate the robustness and versatility of the finite-particle method by means of a test problem concerned with the injection of mass into a sphere. There test results demonstrate the meshless character of the method in any spatial dimension, its ability to redistribute mass particles and follow their evolution in time, its ability to satisfy flux boundary conditions for general domains based solely on a distance function, and its robust convergence characteristics.
\end{abstract}

\maketitle


\section{Introduction}
\label{sec:Introduction}

Particle methods have attained considerable acceptance in solid and fluid mechanics, e.~g, the Smoothed Particle Hydrodynamics method (SPH) \cite{lucy:1977, monaghan:1982}, the Material Point Method \cite{Sulsky94}, the Reproducing Kernel Particle method \cite{liu:1995}, the Corrective Smoothed Particle Method \cite{chen:1999}, the Modified Smoothed Particle Hydrodynamics method \cite{zhang:2004}, the Optimal-Transportation Meshfree method (OTM) \cite{LiHabbalOrtiz2010, Weissenfels:2018}, the dislocation monopole method \cite{Deffo:2019, Ariza:2021}, and others. Whereas the intial development of particle methods was often {\sl ad hoc}, recently there has been increasing awareness of the synergistic connection between particle methods and optimal transport (cf., e.~g., \cite{Evans:1997, Villani2003, Daneri:2014100} for background), in as much as the theory of optimal transport provides a suitable mathematical foundation for formulating particle methods and, conversely, particle methods supply a natural discretization of problems of optimal transport.  

The essential structure of {\sl flow} problems is exemplified by Benamou and Brennier's reformulation of the compressible Euler flow problem \cite{Benamou:2000, LiHabbalOrtiz2010} in terms of two fields, the mass density and the velocity field, which are required to jointly minimize an action integral subject to a continuity equation constraint. In particular, the discretization of the problem requires the formulation of discretization rules for the mass density and for the velocity field. In the OTM method \cite{LiHabbalOrtiz2010, Weissenfels:2018}, the mass density is dicretized into point masses, or {\sl particles}, and the velocity field is discretized by means of conforming interpolation from nodal values. The convergence properties of OTM have been established in \cite{Bompadre2012a, Bompadre2012b, Schmidt:2014} and applications to solid and fluid flows, including plasticity and fragmentation, have been presented in \cite{LiHabbalOrtiz2010, Li:2015, Wessels:2018, Navas:2018, wang2020hot}, among others.

Fedeli {\sl et al.} \cite{fedeli2017geometrically} applied the OTM method to advection-diffusion problems. The connection between diffusion problems and optimal transport was elucidated in the seminal work of Jordan, Kinderlehrer and Otto \cite{JordanKinderlehrerOtto1997, JordanKinderlehrerOtto1998, JordanKinderlehrerOtto1999}. In the approach of Fedeli {\sl et al.} \cite{fedeli2017geometrically} the OTM two-field structure is retained, with the mass density dicretized by means of point masses and the velocity field, which for diffusion problems is entropic in nature, by means of conforming interpolation. A particularly powerful property of optimal transport methods as they bear on advection-diffusion problems is that they are {\sl geometrically exact}, in the sense that all advection and volume updates are computed exactly from the incremental transport mapping. This property is in contrast with upwinding schemes and Eulerian treatments of the volume constraint, which are inexact, costly and prone to instabilities. The geometrical structure of optimal transport (cf., e.~g., \cite{Gangbo:1996, Daneri:2014100} for relevant background) is, in fact, a main source of its computational power. 

In contrast to the OTM two-field representation of diffusion, so-called {\sl blob methods} work solely with the mass density field, which is discretized into smoothed mass particles, or {\sl blobs}, of a certain finite width \cite{carrillo2017, carrillo2019}. For diffusion problems, these finite-particle methods have the appeal of setting forth single-field numerical schemes, at great gain in simplification. However, they also raise a number of numerical analysis questions, such as the consistent variational foundation of the schemes, the enforcement of boundary conditions, the optimal selection of particle thickness and the overall accuracy and convergence of the schemes.

In the present work, we resort to a time-discretized variational formulation of mass diffusion, as in \cite{fedeli2017geometrically}, but discretize the mass density by means of finite particles, or blobs, as in  \cite{carrillo2017, carrillo2019}, in order to formulate a class of velocity-free finite-particle methods. The immediate benefit of using finite-width particles, as opposed to Dirac deltas, is that their entropy can be computed directly from the relative entropy, or Kullback-Leibler, functional (cf., e.~g, \cite{Pinski:2015}). This is in contrast to mass discretization based on point masses, such as used in \cite{fedeli2017geometrically}, for which the entropy functional needs to be carefully rephrased in a weak form. The motion of the finite particles then follows as the result of a competition between entropy, as given by the Kullback-Leibler functional, and mobility, as represented by the Wasserstein distance (cf., e.~g., \cite{Villani2003}) between consecutive configurations. Thus, entropy works to disperse the particles uniformly over the domain, whereas mobility works to hinder their motion.

In Section~\ref{sec:approximation}, we show how this competition results in discrete Euler-Lagrange equations that, in combination with suitable quadrature rules, are explicit in the positions of the particles. A numerical issue that is central to finite-particle schemes such as considered here concerns the efficient choice of particle width. In the present work, we optimize the particle widths {\sl variationally}, i.~e., we determine them by minimizing the same incremental functional that determines the motion of the particles. We show that this optimality criterion results in particle widths on an intermediate scale between the mean-free-distance between particles and the size of the domain, as expected. Another critical issue concerns the formulation of boundary conditions. We show that flux boundary conditions can be conveniently formulated in terms of injection/ejection of particles into/from the domain. We also show that zero-flux conditions can be formulated by means of a barrier potential based solely on a distance function to the domain, which greatly facilitates consideration of general domains defined, e.~g., by means of CAD tools.

In Section~\ref{sec:implementation}, a number of implementational issues are discussed, including the efficient calculation of particle widths and the use of fast searches to accelerate the calculation of pairwise interaction forces. These issues become critical when the number of particles becomes large and need to be addressed carefully in order for the calculations to be feasible. In Section~\ref{sec:NumericalExperiments}, we illustrate the scope and properties of the method by means of a test problem concerned with the injection of mass into a sphere. The ability to inject finite particles into the domain and follow the diffusion-driven trajectories in time, including convergence to a uniform steady-state density after the injection ends, is quite remarkable. The weak convergence of the scheme is monitored through the moment of inertia of the particles at steady state. The test is indicative of robust convergence with respect to the number of particles. 

\section{Problem definition}
\label{sec:ProblemDefinition}

We consider the advection-diffusion initial-boundary-value problem
\begin{subequations}\label{system_complete}
\begin{align}
    &
    \partial_t \rho + \nabla\cdot(\rho u) = \kappa \, \Delta \rho + \rho s,
    & \text{in } \Omega \times [0,T] ,
    \label{eq:TD:Diff1}
    \\ &
    \rho = g,
    & \text{on } \Gamma_D \times [0,T] ,
    \label{eq:TD:Diff2}
    \\ &
    (\kappa \nabla\rho - \rho u )\cdot n = f,
    & \text{on } \Gamma_N \times [0,T] ,
    \label{eq:TD:Diff3}
    \\ &
    \rho(x,0) = \rho_0(x),
    & \text{in } \Omega ,
    \label{eq:TD:Diff4}
\end{align}
\end{subequations}
where $\rho$ is the unknown density or concentration, $\rho_0$ is its initial value, $u$ is a given advection velocity field, $s$ is a source density, $g$ is a prescribed density, $f$ is a prescribed outward mass flux, $\kappa$ is a diffusion coefficient, $\Omega$ is a bounded domain in $\mathbb{R}^d$, $\Gamma = \partial\Omega$ is the boundary of $\Omega$, $n$ is the outward unit normal at the boundary, $\Gamma_D$ is the Dirichlet boundary and $\Gamma_N$ the Neumann boundary, with $\Gamma_D \cap \Gamma_N = \emptyset$ and $\Gamma_D \cup \Gamma_N = \Gamma$.

Problem (\ref{eq:TD:Diff1}) can be equivalently reformulated as the transport problem
\begin{subequations}\label{eq:TD:RV}
\begin{align}
    & \label{eq:TD:RV1}
    \partial_t \rho + \nabla\cdot(\rho v) = \rho s ,
    \\ & \label{rVre5S}
    \rho v = \rho u - \kappa \nabla\rho ,
\end{align}
\end{subequations}
where $v$ is a velocity field that results from the combined effect of advection and diffusion, eq.~(\ref{eq:TD:RV1}) is the continuity equation of Eulerian continuum mechanics and (\ref{rVre5S}) is a mobility law combining the effects of advection and diffusion. The corresponding Lagrangian formulation is
\begin{subequations}
\begin{align}
    & \label{2uN5WW}
    \rho(\varphi(x,t),t)
    =
    \frac{\rho_0(x)}{\det\nabla\varphi(x,t)}
    \exp
    \Big(
        \int_0^t s(\varphi(x,\tau),\tau) \, d\tau
    \Big)
    \\ & \label{yC2krR}
    \partial_t \varphi(x,t)
    =
    v(\varphi(x,t),t) ,
\end{align}
\end{subequations}
where $\varphi : \Omega \times [0,T] \to \Omega$ is the transport map. The mass density
\begin{equation}\label{Meg8nu}
    \rho_0(x)
    \exp
    \Big(
        \int_0^t s(\varphi(x,\tau),\tau) \, d\tau
    \Big)
\end{equation}
in (\ref{2uN5WW}) gives the initial mass density at material point $x$ adjusted for the mass inserted or removed by the sources. Eq.~(\ref{2uN5WW}) states that $\rho(x,t)$ is the push-forward of (\ref{Meg8nu}) by the transport map $\varphi(\cdot,t)$ whereas eq.~(\ref{yC2krR}) relates the transport map to the particle velocity.

Eqs.~(\ref{2uN5WW}) and (\ref{yC2krR}) may be thought as jointly defining an evolution for both the measure $\rho \, dx$ and the transport map $\varphi$. In particular, we note that this extension to measures requires consideration of the transport map $\varphi$ as an additional unknown of the problem.

\section{Approximation}
\label{sec:approximation}

We aim to approximate the solutions of (\ref{system_complete}) variationally through a combination of time discretization followed by spatial discretization of the mass density into particles. To this end, we resort to an incremental variational principle that sets forth an optimality criterion for the approximations. 

\subsection{Fractional steps}

We begin by discretizing time into a sequence of discrete times $t_0, t_1, \dots, t_k, t_{k+1} \dots$ and approximating the corresponding mass densities $\rho_0, \rho_1, \dots, \rho_k, \rho_{k+1} \dots$.
To this end, we exploit the additive structure of (\ref{eq:TD:Diff1}) to decompose incremental updates by the method of fractional steps \cite{fedeli2017geometrically}. The advective fractional step is governed by the pure advection equation
\begin{equation}
    \partial_t \rho + \nabla\cdot(\rho u) = 0 \, ,
\end{equation}
which follows formally from \eqref{eq:TD:RV1} by setting $\kappa = 0$ and $s=0$ and it be conveniently solved exactly following the approach described in \cite{fedeli2017geometrically}. This exact geometrical treatment of advection sets particle methods apart from upwinding methods and confer them great advantage and strength. The source fractional step 
\begin{equation}
    \partial_t \rho = \rho s \, ,
\end{equation}
is likewise amenable to an exact treatment, cf.~(\ref{Meg8nu}). Therefore, in the remainder of the section we focus on the pure diffusional fractional step, obtained by formally dropping the advection and source terms in \eqref{eq:TD:RV1}. Further detail on the convection step and examples of application may be found in \cite{fedeli2017geometrically}. 

\subsection{Variational formulation of the diffusional fractional step}
\label{ssec:VariationalFormulation}

Following \cite{JordanKinderlehrerOtto1997, JordanKinderlehrerOtto1998, JordanKinderlehrerOtto1999, fedeli2017geometrically}, we characterize the incremental evolution of the density by means of the Jordan-Kinderlehrer-Otto (JKO) functional
\begin{equation}\label{eq:TD:F}
    F(\rho_{k+1})
    =
    \frac{1}{2}
    \frac{d_W^2(\rho_k,\rho_{k+1})}{t_{k+1}-t_k}
    +
    \int_\Omega 
        \kappa \rho_{k+1} 
        \log\Big(\frac{\rho_{k+1}}{\rho_\infty}\Big) 
    \, dx
    \to \inf!,
\end{equation}
where $d_W$ is the Wasserstein distance (cf., e.~g., Appendix~A of \cite{fedeli2017geometrically}) and $\rho_\infty$ is a reference density. For definiteness, we specifically choose $\rho_\infty = M/|\Omega|$, where $M$ is the total mass of the system at steady state and $|\Omega|$ is the volume of the domain. 

It is verified that this formulation indeed supplies a time discretization of the mobility law (\ref{rVre5S}) in the limit of zero advection. Thus, a straightforward calculation (cf., e.~g., Appendix~A of \cite{fedeli2017geometrically}) gives the Euler-Lagrange equation
\begin{equation}\label{AD7gzHD3}
    \rho_{k+1}(x)
    \frac{x-\varphi_{k+1\to k}(x)}{t_{k+1}-t_k}
    =
    -
    \kappa \nabla \rho_{k+1}(x) ,
\end{equation}
where $\varphi_{k+1\to k}$ is the incremental transport map, $\varphi_{k\to k+1} = \varphi_{k+1\to k}^{-1}$ and $\rho_{k+1}$ follows from $\varphi_{k+1\to k}$ through the geometrically exact pushforward operation
\begin{equation}\label{fAVOs7}
    \rho_{k+1} \circ \varphi_{k\to k+1}
    =
    \rho_k/\det\big(\nabla\varphi_{k\to k+1}\big) .
\end{equation}
We verify that (\ref{AD7gzHD3}) is indeed a time discretization of (\ref{rVre5S}). A rigorous analysis of convergence with respect to the time step may be found in \cite{JordanKinderlehrerOtto1997, JordanKinderlehrerOtto1998}. 

\subsection{Spatial discretization}
\label{ssec:SpatialDiscretization}

From the standpoint of weak convergence of measures, 
it would be natural to approximate $\rho(x,t)$ by concentrating mass on a finite collection of points $\{x_p(t)\}_{p=1}^n$, i.~e., by setting
\begin{equation}\label{8EYqqq}
    \rho(x,t) \sim \sum_{p=1}^n m_p \delta(x-x_p(t)) ,
\end{equation}
where $\delta$ is the Dirac delta and $m_p$ is the mass carried by particle $p$. However, the functional (\ref{eq:TD:F}) is not defined for trial densities of this type. In \cite{fedeli2017geometrically}, this difficulty is circumvented by replacing the incremental functional (\ref{eq:TD:F}) with an alternative weak statement, modeled after (\ref{eq:TD:RV}), in which the mass density appears linearly and can therefore be approximated by the {\sl ansatz} (\ref{8EYqqq}).

In the present work, following Carrillo {\sl et al.} \cite{carrillo2017, carrillo2019} we instead spread, or 'fatten', the particles over a finite width, to be determined. The corresponding approximation is
\begin{equation}\label{bC4KRV}
    \rho(x,t)
    \sim
    \sum_{p=1}^n m_p  \varphi_p(x - x_p(t), t) ,
\end{equation}
where the function $\varphi_p(x - x_p(t); t)$ represents the time-dependent profile of particle $x_p$. The form of the profile $\varphi_p(x - x_p(t), t)$ may be parametrized within a convenient class of functions and the explicit dependence on $t$ then represents the time-dependence of the parameters. The discrete system is then obtained by inserting the {\sl ansatz} (\ref{bC4KRV}) into the functional (\ref{eq:TD:F}) and rendering the resulting discrete functional stationary with respect to all degrees of freedom.

\subsection{Gaussian particle profile}

We specifically choose a Gaussian particle profile of the form
\begin{equation}\label{eq:nfunction}
    \varphi_p(x - x_p(t), t)
    =
    \Big( \frac{\beta_p(t)}{\pi} \Big)^{d/2}
    \, {\rm e}^{-\beta_p(t) \| x - x_p(t) \|^2}
    :=
    {N}(x - x_p(t), \beta_p(t)),
\end{equation}
normalized so that
\begin{equation}
    \int_{\mathbb{R}^d} {N}(x - x_p(t), \beta_p(t)) \, dx = 1 .
\end{equation}
We note that ${N}(x - x_p(t), \beta_p(t))$ has dimensions of an inverse volume, carries unit mass and has a width set by the parameter $\beta_p(t)$. 

\subsection{Wasserstein distance}

The Gaussian profile of the particles and the absence of mass redistribution renders the calculation of the Wasserstein distance $d_W^2(\rho_k,\rho_{k+1})$ in (\ref{eq:TD:F}) straightforward. Thus, for particles of constant mass $d_W^2(\rho_k,\rho_{k+1})$ simply records the total cost of transportation of all the particles from their initial to their final positions, namely,
\begin{equation}
    d_W^2(\rho_k,\rho_{k+1})
    \sim 
    \sum_{p=1}^n 
        d_W^2\big(m_p N_{p,k}(x), m_p N_{p,k+1}\big) ,
\end{equation}
where we write $N_{p,k}(x)$ $:=$ ${N}(x - x_{p,k}, \beta_{p,k})$, $N_{p,k+1}(x)$ $:=$ ${N}(x - x_{p,k+1}, \beta_{p,k+1})$, and neglect the effect of the clipping of the tails of the Gaussians by a finite domain. Conveniently, the Wasserstein distance between Gaussian measures is known explicitly \cite{Dowson:1982, Olkin:1982, Givens:1984, Knott:1984}, whence we obtain
\begin{equation}
    d_W^2\big(m_p N_{p,k}(x), m_p N_{p,k+1}\big)
    \sim 
    \Big(
        \| x_{p,k+1} - x_{p,k} \|^2
        +
        | \beta_{p,k+1}^{-1/2} - \beta_{p,k}^{-1/2} |^2
    \Big)
    \, m_p .
\end{equation}
We note that $d_W^2(\rho_k,\rho_{k+1})$ quantifies not only the extent of particle displacement during a time step but also the change in the width of the particles, if any.

\subsection{Hermitian quadrature}

The choice of Gaussian profiles also facilitates the computation of integrals by numerical quadrature. Thus, for densities of the form (\ref{bC4KRV}) and (\ref{eq:nfunction}) we have
\begin{equation} 
\begin{split}
    &
    \int_\Omega \kappa \rho_{k+1} \log\rho_{k+1} \, dx
    = 
    \sum_{p=1}^n
    \int_\Omega \kappa \log\Big(\frac{\rho_{k+1}(x)}{\rho_\infty}\Big) 
        \times \\ & \qquad\qquad\qquad
        \Big( \frac{\beta_{p,k+1}}{\pi} \Big)^{d/2}
        \exp\Big( -\beta_{p,k+1} \| x - x_{p,k+1} \|^2 \Big) 
        \, m_p
    \, dx .
\end{split}
\end{equation}
The Gaussian weight in the integral suggests approximation by Hermitian quadrature \cite{hildebrand2013introduction}. The one-point quadrature rule is particularly simple and takes the form
\begin{equation}\label{P3AvN2}
    \int_\Omega \kappa \rho_{k+1}(x) \log\Big(\frac{\rho_{k+1}(x)}{\rho_\infty}\Big) \, dx
    \sim 
    \sum_{p=1}^n
    \kappa \log\Big(\frac{\rho_{k+1}(x_{p,k+1})}{\rho_\infty}\Big)
    \, m_p .
\end{equation}
Higher accuracy can be achieved by recourse to quadrature rules of higher order, albeit at increased computational cost and complexity.

\subsection{The discrete problem}

Combining the preceding approximations, the reduced JKO functional takes the form
\begin{equation}\label{ZXTj1Y}
\begin{split}
    F(\rho_{k+1})
    & \sim
    \sum_{p=1}^n
    \frac{1}{2}
    \frac
    {
        \| x_{p,k+1} - x_{p,k} \|^2 + ( \beta_{p,k+1}^{-1/2} - \beta_{p,k}^{-1/2} )^2
    }
    {
        t_{k+1}-t_k
    } 
    \, m_p
    \\ & +
    \sum_{p=1}^n
    \kappa
    \log
    \Big(
        \frac{\rho_{k+1}(x_{p,k+1})}{\rho_\infty}
    \Big)
    \, m_p
    \to \min!
\end{split}
\end{equation}
with
\begin{equation}\label{2IeJn6}
    \rho_{k+1}(x)
    :=
    \sum_{p=1}^n
    \Big( \frac{\beta_{p,k+1}}{\pi} \Big)^{d/2}
    \,{\rm e}^{- \beta_{p,k+1} \| x - x_{p, k+1} \|^2 }
    \, m_p .
\end{equation}
We note that the incremental variational principle (\ref{ZXTj1Y}) determines both the updated position of the particles and their optimal widths. The corresponding Euler-Lagrange equations are
\begin{equation}\label{9eBxCb}
    \frac{\partial F}{\partial x_{r,k+1}}
    =
    0 ,
    \qquad
    \frac{\partial F}{\partial \beta_{r,k+1}}
    =
    0 ,
\end{equation}
which evaluate to
\begin{subequations}\label{j4K0Ht}
\begin{align}
    & \label{bXq5IE}
    m_r
    \frac{x_{r,k+1} - x_{r,k}}{t_{k+1}-t_k}
    +
    \frac{\partial}{\partial x_{r,k+1}}
    \sum_{p=1}^n
    \kappa
    \log
        \Big(
            \frac{\rho_{k+1}(x_{p,k+1})}{\rho_\infty}
        \Big)
    \, m_p
    =
    0 ,
    \\ & \label{W0gjO2}
    -
    \frac{m_r}{2}
    \frac
    {
        \beta_{r,k+1}^{-1/2}-\beta_{r,k}^{-1/2}
    }
    {
        \beta_{r,k+1}^{3/2} (t_{k+1} - t_k)
    }
    +
    \frac{\partial}{\partial \beta_{r,k+1}}
    \sum_{p=1}^n
        \kappa
        \log
        \Big(
            \frac{\rho_{k+1}(x_{p,k+1})}{\rho_\infty}
        \Big)
    \, m_p
    =
    0 .
\end{align}
\end{subequations}
We note that these equations represent an implicit forward-Euler update for the positions and widths of the particles. 

\subsection{Boundary conditions}

Neumann, or flux, boundary conditions can be readily implemented within the particle representation just outlined. Thus, a prescribed inward flux can be built into the calculations by injecting particles through the boundary of the domain so as to achieve the desired mass input rate per unit area. A prescribed outward flux can be implemented either by removing particles crossing the boundary, or by inserting particles with negative mass \cite{Profeta:2020}, or a combination of both, again so as to achieve the desired mass input rate per unit area.

A zero flux condition can be implemented by means of a barrier potential, i.~e., a continuous differentiable function $V(x)$ that is zero on the domain, including its boundary, and grows steeply away from it. For instance, in calculations we choose
\begin{equation}\label{Pe18A9}
    V(x) = \frac{C}{2} {\rm dist}^2(x,\Omega) ,
\end{equation}
where
\begin{equation}
    {\rm dist}(x,\Omega)
    =
    \min\{ \|x-y\| \, : \, y \in \Omega \}
\end{equation}
is the distance function to $\Omega$ and the constant $C$ is to be calibrated. The incremental problem (\ref{ZXTj1Y}) is then augmented to
\begin{equation}
    F(\rho_{k+1})
    +
    \sum_{p=1}^n V(x_p) \, m_p
    \to
    \min!
\end{equation}
Evidently, the effect of the barrier potential is to apply a restoring diffusional force $DV(x_p) \, m_p$ to particles $x_p$ that wander out of the domain, causing them to return to it. 

Conveniently, the distance function $d(x,\Omega)$ is a functionality that is provided by most Computer-Aided Design (CAD) codes, which facilitates the implementation of the finite-particle method based on CAD models of the domain of analysis.

A number of extensions of optimal transport and the JKO functional have been proposed for problems with Dirichlet boundary conditions that can be applied to particle methods \cite{Figalli:2010, Dweik:2018a, Dweik:2018b, Profeta:2020}. A possible simple implementation of Dirichlet boundary conditions consists of fixing particles on the Dirichlet boundary at the desired areal mass density. However, such extensions are beyond the scope of this paper and will not be addressed further.

\section{Numerical implementation}
\label{sec:implementation}

In this section, we discuss ways in which the discrete problem (\ref{j4K0Ht}) can be simplified, and the solution accelerated, without significantly compromising accuracy or convergence. 

\subsection{Computation of diffusive forces}

It is possible to derive a particular expression for the diffusive forces in (\ref{bXq5IE}) within the accuracy of the Hermitian quadrature rule. To this end, we write
\begin{equation}\label{1oZA7P}
\begin{split}
    &
    \frac{\partial}{\partial x_{r,k+1}}
    \sum_{p=1}^n
    \kappa
    \log
        \Big(
            \frac{\rho_{k+1}(x_{p,k+1})}{\rho_\infty}
        \Big)
    \, m_p
    = \\&
    \sum_{p=1}^n
    \kappa
    \Big[
        \frac{\partial}{\partial x_{r,k+1}}
        \log
        \Big(
            \frac{\rho_{k+1}(x)}{\rho_\infty}
        \Big)
    \Big]_{x=x_{p,k+1}}
    \, m_p
    + 
    \kappa \nabla \rho_{k+1}(x_{r,k+1}) m_r ,
\end{split}
\end{equation}
with $\rho_{k+1}(x)$ as in (\ref{2IeJn6}).  Within the accuracy of one-point Hermitian interpolation, we have
\begin{equation}
\begin{split}
    &
    \sum_{p=1}^n
    \kappa
    \Big[
        \frac{\partial}{\partial x_{r,k+1}}
        \log
        \Big(
            \frac{\rho_{k+1}(x)}{\rho_\infty}
        \Big)
    \Big]_{x=x_{p,k+1}}
    \, m_p
    = \\ &
    \sum_{p=1}^n
    \kappa
    \Big[
        \frac{1}{\rho_{k+1}(x)}
        \frac{\partial}{\partial x_{r,k+1}}\rho_{k+1}(x)
    \Big]_{x=x_{p,k+1}}
    \, m_p
    \sim \\ &
    \int_\Omega
        \kappa
        \frac{\partial}{\partial x_{r,k+1}}\rho_{k+1}(x)
    \, dx
    = 
    \frac{\partial}{\partial x_{r,k+1}}
    \int_\Omega
        \kappa
        \rho_{k+1}(x)
    \, dx
    \sim
    0 .
\end{split}
\end{equation}
To within this approximation, (\ref{1oZA7P}) therefore simplifies to the particularly simple expression
\begin{equation}\label{zN9kBU}
\begin{split}
    &
    \frac{\partial}{\partial x_{r,k+1}}
    \sum_{p=1}^n
    \kappa
    \log
        \Big(
            \frac{\rho_{k+1}(x_{p,k+1})}{\rho_\infty}
        \Big)
    \, m_p
    \sim 
    \kappa \nabla \rho_{k+1}(x_{r,k+1}) \, m_r ,
\end{split}
\end{equation}
which we use in all calculations.

\subsection{Uniform particle width}

Considerable simplification is achieved if the width of the particles is assumed to be uniform, i.~e., $\beta_{r,k+1} = \beta_{k+1}$. Inserting this assumption into (\ref{ZXTj1Y}) and rendering the functional stationary with respect to $\beta_{k+1}$, (\ref{W0gjO2}) simplifies to the single equation
\begin{equation}\label{SjT874}
    -
    \frac{M}{2}
    \frac
    {
        \beta_{k+1}^{-1/2}-\beta_{k}^{-1/2}
    }
    {
        \beta_{k+1}^{3/2} (t_{k+1} - t_k)
    }
    +
    \frac{\partial}{\partial \beta_{k+1}}
    \sum_{p=1}^n
        \kappa
        \log
        \Big(
            \frac{\rho_{k+1}(x_{p,k+1})}{\rho_\infty}
        \Big)
    \, m_p
    =
    0 ,
\end{equation}
where $M=\sum_{r=1}^n m_r$ is the total mass of the system and
\begin{equation}\label{Tw3L9Z}
    \rho_{k+1}(x)
    :=
    \sum_{p=1}^n
        \Big( \frac{\beta_{k+1}}{\pi} \Big)^{d/2}
        \,{\rm e}^{- \beta_{k+1} \| x - x_{p,k+1} \|^2 }
    \, m_p .
\end{equation}
in place of (\ref{2IeJn6}). Evaluating the derivative in (\ref{SjT874}) using (\ref{Tw3L9Z}) yields
\begin{equation} \label{a6rU9b}
\begin{split}
    &
    \frac{\partial}{\partial \beta_{k+1}}
    \sum_{p=1}^n
        \kappa
        \log
        \Big(
            \frac{\rho_{k+1}(x_{p,k+1})}{\rho_\infty}
        \Big)
    \, m_p
    = 
    M \kappa
    \frac{d}{2}\frac{1}{\beta_{k+1}}
    - \\ &
    \sum_{p=1}^n
        \kappa
        \Big(
            \frac
            {
                \sum_{q=1}^n
                    \| x_{p,k+1} - x_{q,k+1} \|^2
                    \,{\rm e}^{- \beta_{k+1} \| x_{p,k+1} - x_{q,k+1} \|^2 }
                \, m_q 
            }
            {
                \sum_{q=1}^n
                    \,{\rm e}^{- \beta_{k+1} \| x_{p,k+1} - x_{q,k+1} \|^2 }
                \, m_q 
            }
        \Big)
    \, m_p .
\end{split}
\end{equation}

A further simplification is obtained if we assume that the particle widths are at steady state at all times. Under this assumption, (\ref{a6rU9b}) further reduces to the algebraic equation
\begin{equation}\label{L3sNc4}
\begin{split}
    &
    M 
    \frac{d}{2}\frac{1}{\beta_{k+1}}
    = \\ &
    \sum_{p=1}^n
        \Big(
            \frac
            {
                \sum_{q=1}^n
                    \| x_{p,k+1} - x_{q,k+1} \|^2
                    \,{\rm e}^{- \beta_{k+1} \| x_{p,k+1} - x_{q,k+1} \|^2 }
                \, m_q 
            }
            {
                \sum_{q=1}^n
                    \,{\rm e}^{- \beta_{k+1} \| x_{p,k+1} - x_{q,k+1} \|^2 }
                \, m_q 
            }
        \Big)
    \, m_p ,
\end{split}
\end{equation}
which gives the instantaneous width of the particles as a function of their positions. 

\subsection{Optimal particle width at steady state}

Suppose that, at steady state, the mass density attains the constant value $\rho_\infty = M/|\Omega|$, where $M$ is the total mass and $|\Omega|$ the volume of the domain, assumed to be finite. Suppose further that this steady state is approximated by a collection of $n$ particles of uniform width, 
\begin{equation}
    \rho(x)
    :=
    \sum_{p=1}^n
    \Big( \frac{\beta}{\pi} \Big)^{d/2}
    \, {\rm e}^{- \beta \| x - x_{p} \|^2 }
    \, m_p ,
\end{equation}
as in (\ref{2IeJn6}). The corresponding relative entropy then follows as
\begin{equation}\label{T1hz8l}
    F_\infty(\beta)
    =
    \int_\Omega 
        \kappa \rho(x)
        \log\Big(\frac{\rho(x)}{\rho_\infty}\Big) 
    \, dx ,
\end{equation}
to be minimized with respect to $\beta$. To this end, we rewrite (\ref{T1hz8l}) as
\begin{equation}\label{EKon4i}
    F_\infty(\beta)
    =
    \int_{\mathbb{R}^d}
        \kappa \rho(x)
        \log\Big(\frac{\rho(x)}{\rho_\infty}\Big) 
    \, dx 
    -
    \int_{\mathbb{R}^d\backslash\Omega}
        \kappa \rho(x)
        \log\Big(\frac{\rho(x)}{\rho_\infty}\Big) 
    \, dx , 
\end{equation}
and proceed to estimate each term in turn. We note that in the first term in (\ref{EKon4i}) the tails of the finite particles are allowed to extend outside the domain, as tacitly assumed in the quadrature rule (\ref{P3AvN2}). The second term in (\ref{EKon4i}) then corrects for the incurred mass leakoff. 

For large $\beta$ the leakoff mass is confined to a narrow layer of thickness $1/\sqrt{\beta}$ in the vicinity of the boundary $\partial\Omega$ of the domain. For sufficiently regular domains, therefore,
\begin{equation}
    \int_{\mathbb{R}^d\backslash\Omega}
        \kappa \rho(x)
        \log\Big(\frac{\rho(x)}{\rho_\infty}\Big) 
    \, dx 
    \sim
    -
    \kappa \rho_\infty |\partial\Omega| \beta^{-1/2} ,
\end{equation}
where $|\partial\Omega|$ is the area of the domain boundary. We note that the effect of mass leakoff vanishes in the limit of $\beta \to +\infty$, as expected. To estimate the first term of (\ref{EKon4i}), we partition the domain $\Omega$ into $n$ subdomains of volume $|\Omega|/n$, each containing one particle. Again, for sufficiently large $\beta$, we have $\log(\rho(x)) \sim - \beta \|x-x_p\|^2$, in the area containing particle $x_p$, and we compute
\begin{equation}
    \int_{\mathbb{R}^d}
        \kappa \rho(x)
        \log\Big(\frac{\rho(x)}{\rho_\infty}\Big) 
    \, dx 
    \sim
    -
    \kappa \rho_\infty |\Omega| \beta \Big( \frac{|\Omega|}{n} \Big)^{2/d} .
\end{equation}
Inserting these estimates into (\ref{EKon4i}) and minimizing, we obtain
\begin{equation}\label{t80S3G}
    \beta
    \sim
    \Big(
        \frac{|\partial\Omega|}{|\Omega|^{(d+2)/d}}
    \Big)^{2/3}
    \, N^{4/3d} ,
\end{equation}
modulo a multiplicative constant, to be calibrated. In (\ref{t80S3G}), the first factor is of the form $R^{-2}$, where $R$ is an effective radius of the domain. The second factor defines a precise scaling $\beta^{-1/2} \sim N^{-2/3d}$ for the particle width with respect to the number of particles. Remarkably, the optimal particle size takes values on a scale intermediate between the size of the particle neighborhoods, $(|\Omega|/n)^{1/d}$ and the size of the domain $|\Omega|^{1/d}$. 

\subsection{Local-neighborhood searches}
\label{ssec:searches}

A na\"ive evaluation of the diffusive forces (\ref{zN9kBU}) is an $O(n^2)$ operation which, for large $n$, becomes the main computational bottleneck. The complexity of the operation can be reduced---and the calculations accelerated---by noting that, due to the Gaussian decay of the particle profiles, points $x_{p,k+1}$ in the sum (\ref{2IeJn6}) for the calculation of $\nabla \rho_{k+1}(x_{r,k+1})$ at distances much greater than $1/\sqrt{\beta_{r,k+1}}$ from the $x_{r,k+1}$ contribute neglibly to the sum. Therefore, the sum can be restricted to a local neighborhood of $x_{r,k+1}$ on the scale of $1/\sqrt{\beta_{r,k+1}}$ without appreciable loss of accuracy. 

The rate-limiting step in this approach is the operation of searching for the local neighborhoods as they evolve during in calculations as a result of the relative shuffling of the particles. Numerous techniques have been developed for solving the local-neighborhood problem. One approach makes use of space partitioning algorithms where the branch and bound method \cite{Land:2010} is applied. Early examples in this class are the $k$-d tree \cite{Bentley:1975, Friedman:1977} and the $k$-means \cite{Fukunaga:1975} algorithms, with many variants thereof. A deficiency of exact methods is that they cannot guarantee logarithmic search time \cite{Uhlmann:1991, Sproull:1991}. Conveniently, for many applications approximate nearest-neighbor (ANN) algorithms \cite{Muja:2009, Muja:2014}, which trade off accuracy for efficiency, suffice. More recent work has focused on proximity graph-based methods \cite{Li:2019, Groh:2022}, which are well-suited to high-dimensionality problems arising in image recognition, computational linguistics, product recommendation, and others. 

\begin{figure}
    \centering
    \includegraphics[width = 0.6\textwidth]{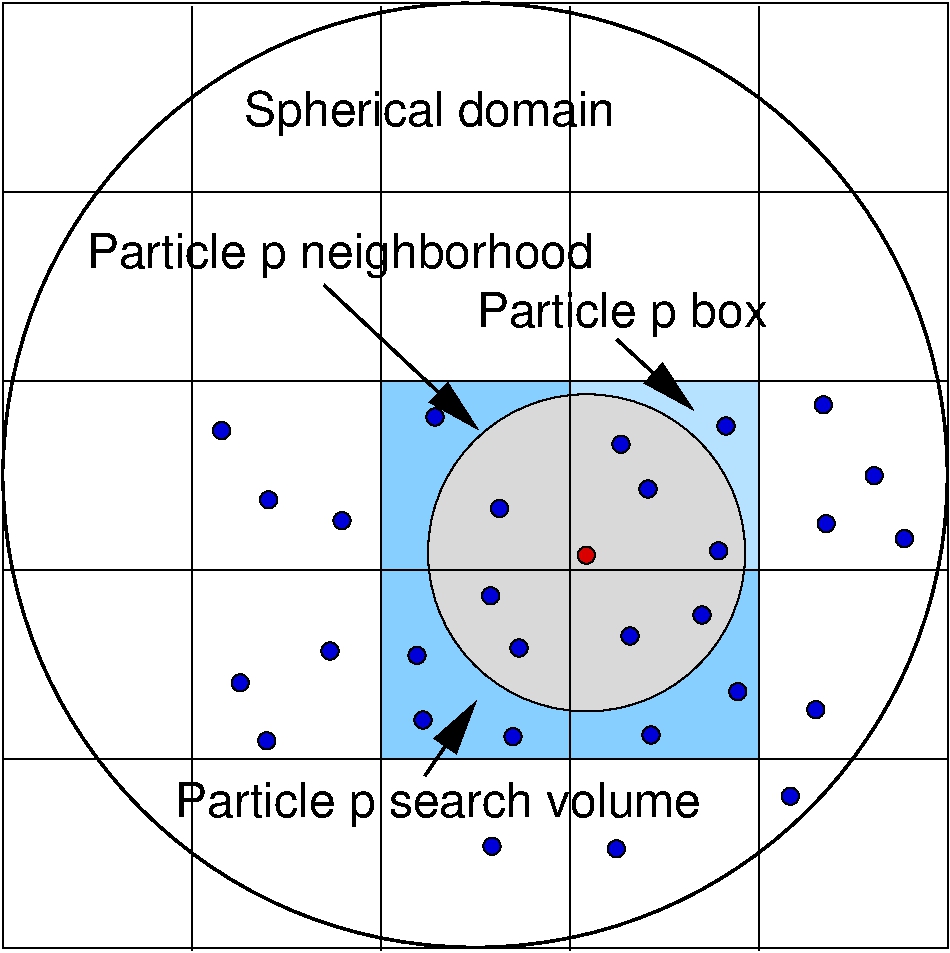}
    \caption{\footnotesize Schematic of the search algorithm with reference to the problem of an injected sphere. The spherical domain is subdivided in same-size cubic cells, or boxes, and each particle is assigned to a cell according to its current position. A cell-connectivity array lists the particles contained in each cell. For each particle, a second nearest-neighbor array records the particles contained within a spherical neighborhood of radius $\sim 1/\sqrt{\beta}$. At every time step, each particle interacts only with the particles in its neighborhood. Periodically, the nearest-neighbor array is updated by means of searches restricted to the cells that intersect the neighborhood of each particle.}
    \label{Fig:SearchAlgorithm}
\end{figure}

The best choice of search algorithm depends on data-related properties such as dimension, data-set size, correlations, density distribution, and others, and performance metrics such as query time, accuracy, query workload, building time and memory usage. In this work, we employ a simple {\sl ad hoc} algorithm based on a regular one-level subdivision of the computational domain into cubic cells, or boxes, Fig.~\ref{Fig:SearchAlgorithm}. For simplicity, we consider the case of uniform and constant $\beta$ pegged, e.~g., to the estimate (\ref{t80S3G}). In this case, the size of cells is taken as a multiple of $1/\sqrt{\beta}$, with factors to be calibrated numerically. Thus, the algorithm requires two connectivity arrays. At initialization, as well as for any new particle entering the domain as a result of a prescribed flux, the particles are assigned to a specific cell according to their coordinates. Each particle is outfitted with the connectivity list of interacting particles, and each cell is outfitted with the list of the particles contained in the cell. In shared memory platforms, the arrays may be conveniently implemented as C++ STL containers. The algorithm keeps the cell array up-to-date by transferring the particles between cells at every time step as needed. By contrast, the particle connectivity lists are updated only after a prescribed number of time steps. The update is carried out by searching only the particles contained in the cells intersected by the spherical neighborhood of the query particle. 

\section{Example: Mass injection into a sphere}
\label{sec:NumericalExperiments}

We demonstrate the properties of the finite-particle scheme defined in the foregoing by means of a test case concerned with the injection of mass into a spherical domain through a small orifice in its surface. 

\subsection{Problem definition}

We consider a spherical domain of radius $R$, initially empty, which is filled with mass injected through a small orifice of area $A$ at the North pole at constant mass flux $f$ during an injection time interval $[0,T_0]$. Subsequently, the particles diffuse freely, eventually filling the domain and attaining a uniform steady-state density $\rho_\infty = M/|\Omega|$. The total injected mass $M = f A T_0$ is divided into $n$ particles of equal mass $m=M/n$, each inserted at time intervals of $T_0/n$ to a position below the North pole offset by a distance $d$ from the boundary. The velocity of injection of the particles is, therefore, $v =d/(T_0/n)$. The problem data (dimensionless units) are collected in table~\ref{table:problemData}. To analyze the performance and the convergence properties of the method, we consider a total diffusion time $T > T_0$ and numbers of particles ranging from a minimum of $125$ to a maximum of $128000$. The parameters used in calculations are listed in Table~\ref{table:problemData}.

\begin{table}[!h]
    \centering
    \begin{tabular}{cccccc}
    \hline
        $R$ & $M$ & $\kappa$ & $T$  & $T_0$ & $v$ \\
        \hline
        1 & 200 & 1 & 120 & 25 & 1 \\
        \hline
    \end{tabular}
    \caption{Parameters used in the sphere-injection test calculations.}
    \label{table:problemData}
\end{table}

The discrete incremental problem (\ref{j4K0Ht}) is solved by a sequential Runge-Kutta four-point scheme (RK4) for small systems, $n \leq 8,000$, and a parallel implementation of forward-Euler on $1$ to $32$ cores for larger systems. The diffusional forces are computed as in (\ref{zN9kBU}). For simplicity, the particle width $1/\sqrt{\beta}$ is assumed to be equal for all particles and constant in time, with value given by the scaling relation (\ref{t80S3G}), calibrated as $\beta = 10 \, n^{4/9}/R^2$, as function of the number of particles. The subdivision scheme described in Section~\ref{ssec:searches} is then used with tolerance $0.01$ to accelerate the calculations. 

Owing to the explicit nature of the calculations, the stable time step is constrained by the diffusional CFL condition $\Delta t \sim \Delta x^2/\kappa$, where we take $\Delta x \sim (|\Omega|/n)^{1/d}$ as an estimate of the distance between particles. Conveniently, the stable time step is always smaller than the injection time step. We additionally enforce the zero-flux boundary condition on the surface of the sphere by means of the barrier potential (\ref{Pe18A9}), with the highest constant $C$ that does not decrease the stable time step. 

\subsection{Particle flow visualizations}
\label{ssec:Results}

\begin{figure}[!h]
    \centering
    \subfigure[$n$=4,000]
    {\includegraphics[width = 0.45\textwidth]{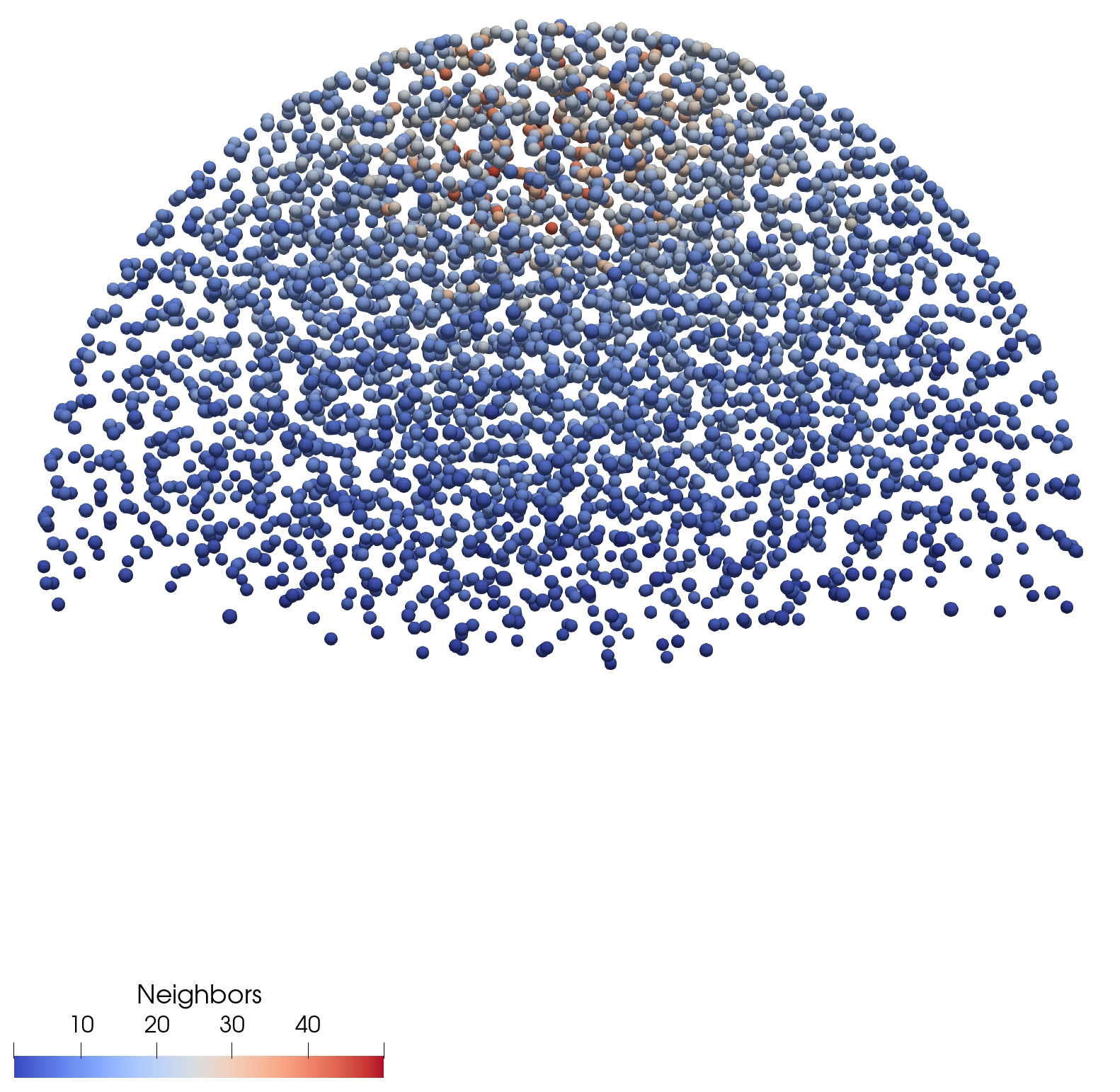}}
    \subfigure[$n$=8,000]
    {\includegraphics[width = 0.45\textwidth]{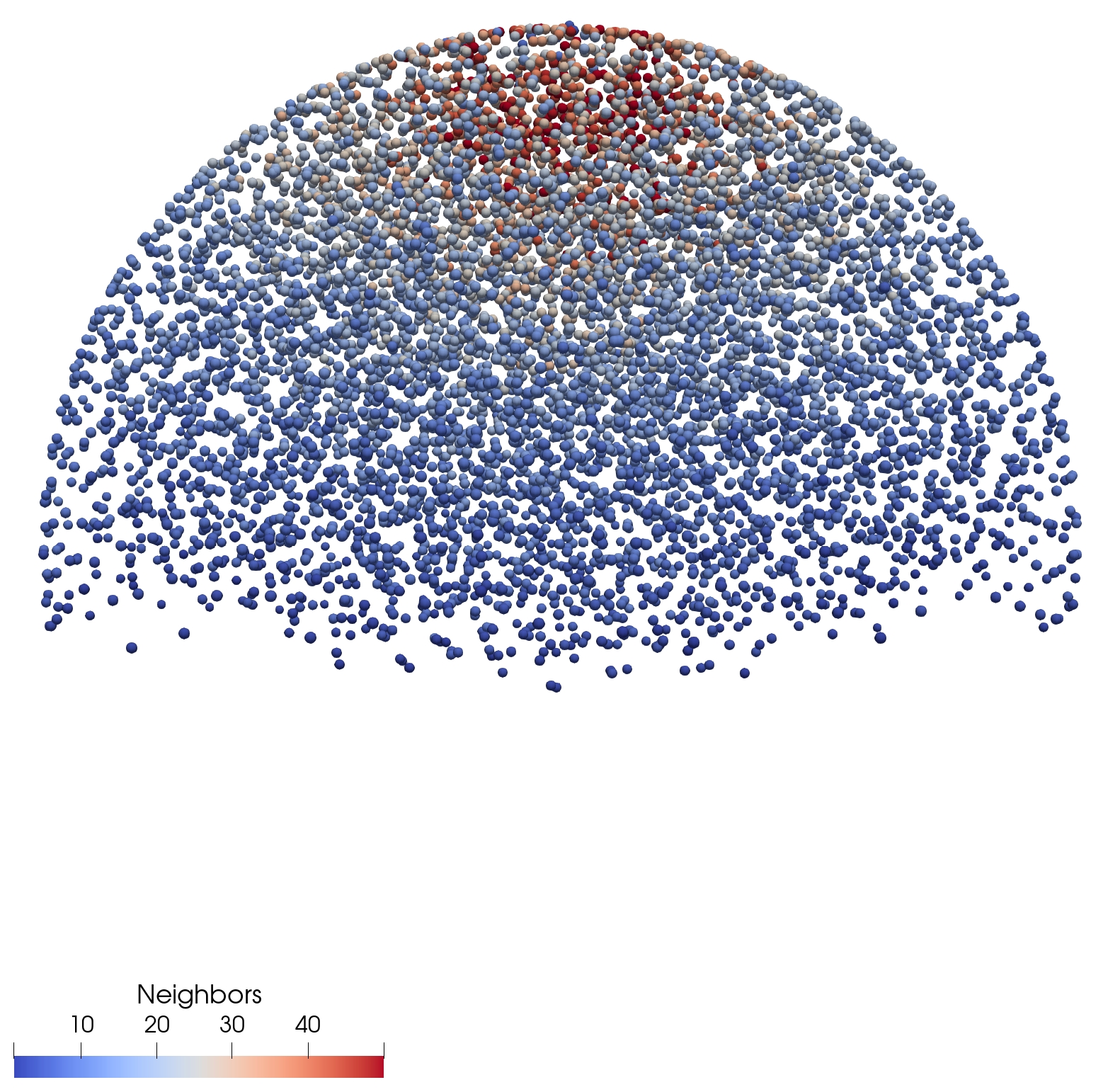}}
    \subfigure[$n$=16,000]
    {\includegraphics[width = 0.45\textwidth]{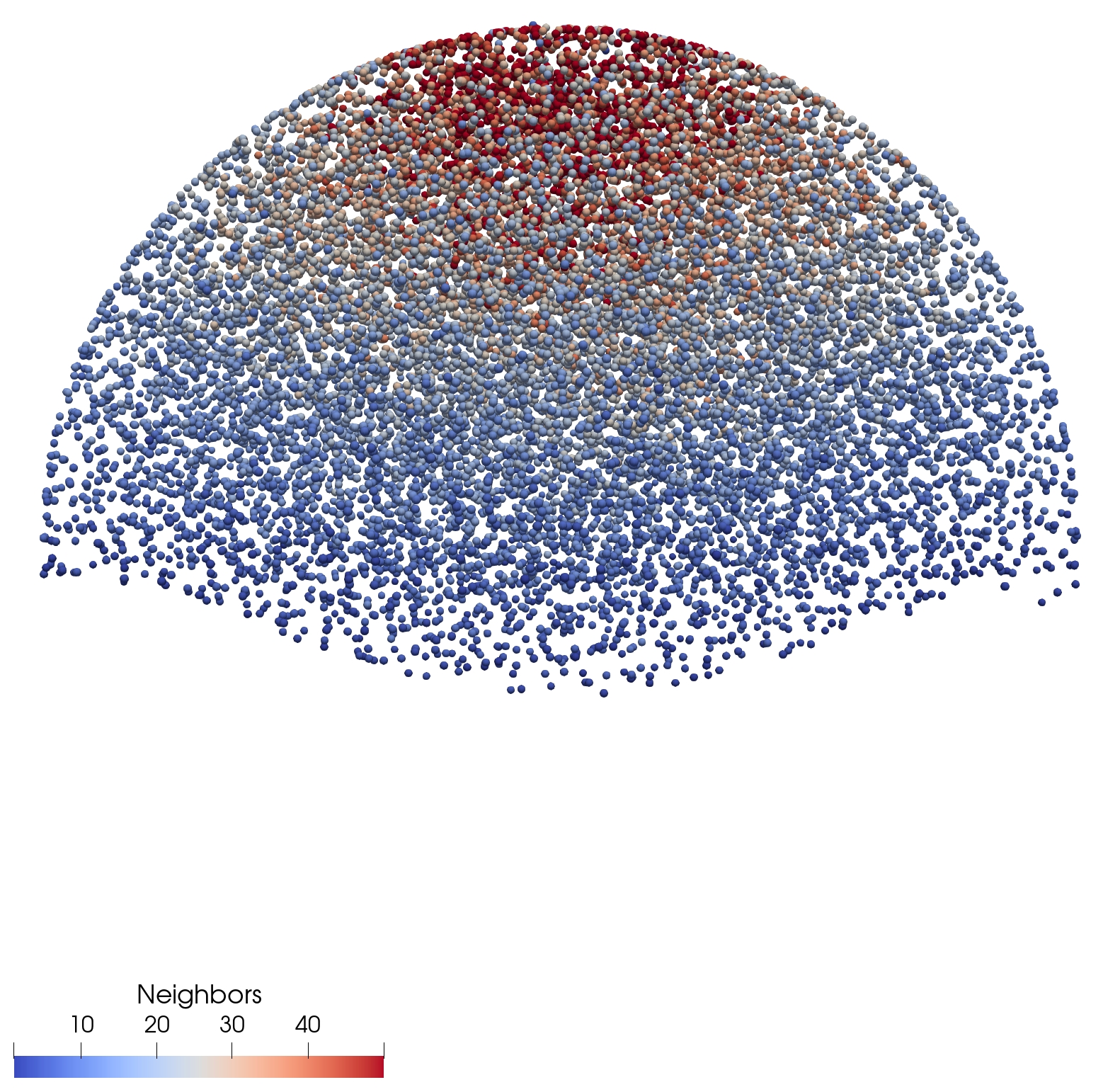}}
    \subfigure[$n$=32,000]
    {\includegraphics[width = 0.45\textwidth]{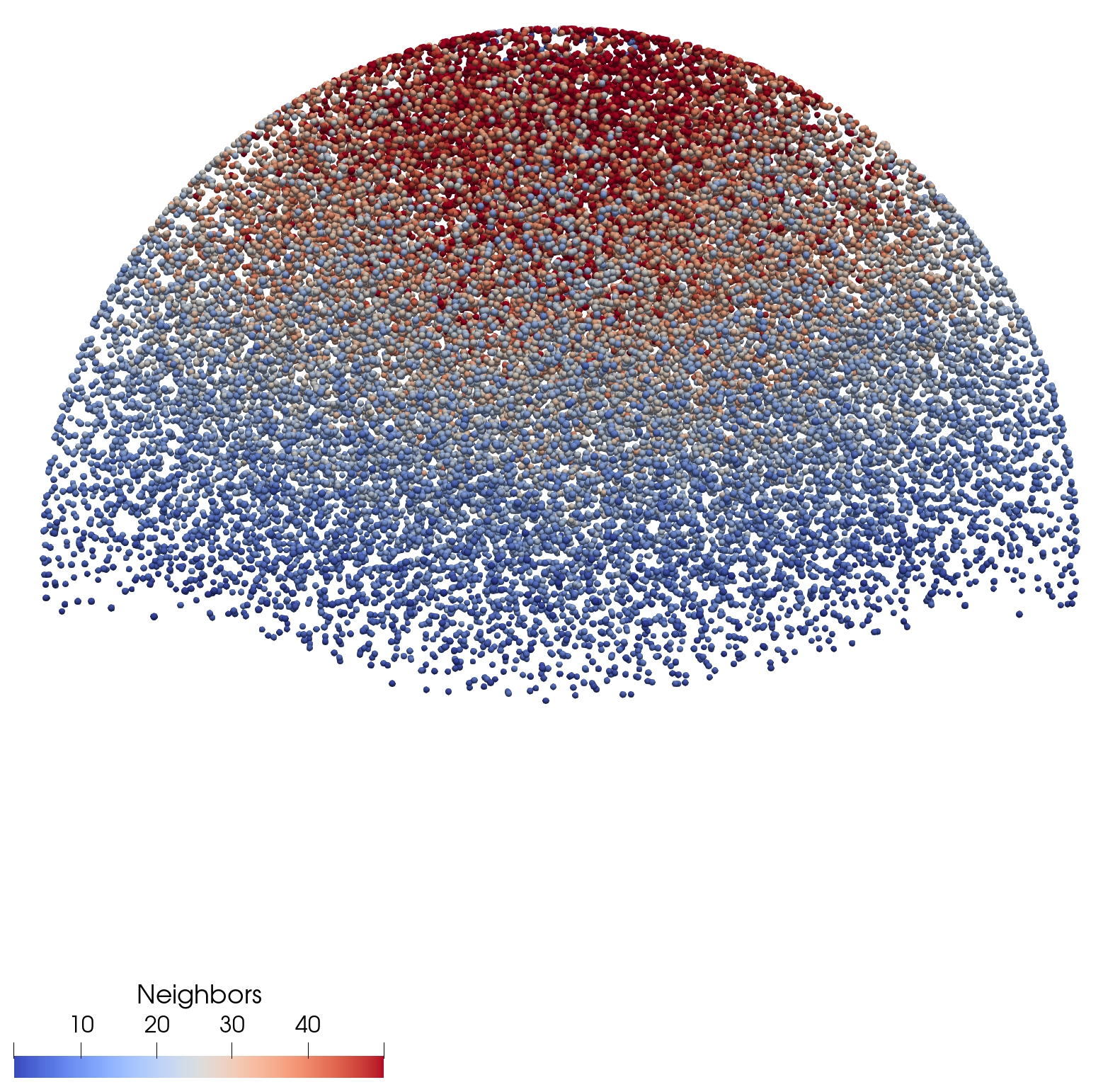}}
    \caption{\footnotesize Visualization of the 3D distribution of the centers of the particles at the end of the injection phase $T_0$ for different number of particles. The analysis adopts a constant value of $\beta$ pegged through the optimal scaling relation to the number of particles $n$. The color of each particle relates to the number of particles in its neighborhood. The size of the particles is related to their mass $m=M/n$.}
    \label{Fig:injectionEnd}
\end{figure}

The evolution of the particle distribution during and after injection is visualized by a sequence of successive snapshots. Fig.~\ref{Fig:injectionEnd} shows the three-dimensional distributions of the centers of the particles at the end of the injection time $T_0$ for four different values of $n$. Every particle is colored according to the number of particles contained in its $1/\sqrt{\beta}$-neighborhood and its size is related to the actual mass $m$. The distribution of particles is characterized by short-range disorder, or randomness, and long-range order, or smoothness, and a general trend towards convergence is evident from the figures. As dictated by optimal scaling, the particle width increases sublinearly with the number of particles and the size of the particle neighborhoods also increases concomitantly, up to a value in the order of $50$ particles for $n=32,000$. The value of the variational principle in elucidating these intricate tradeoffs is remarkable. 

\begin{figure}[!h]
    \centering
    \subfigure[$n$=4,000]
    {\includegraphics[width = 0.45\textwidth]{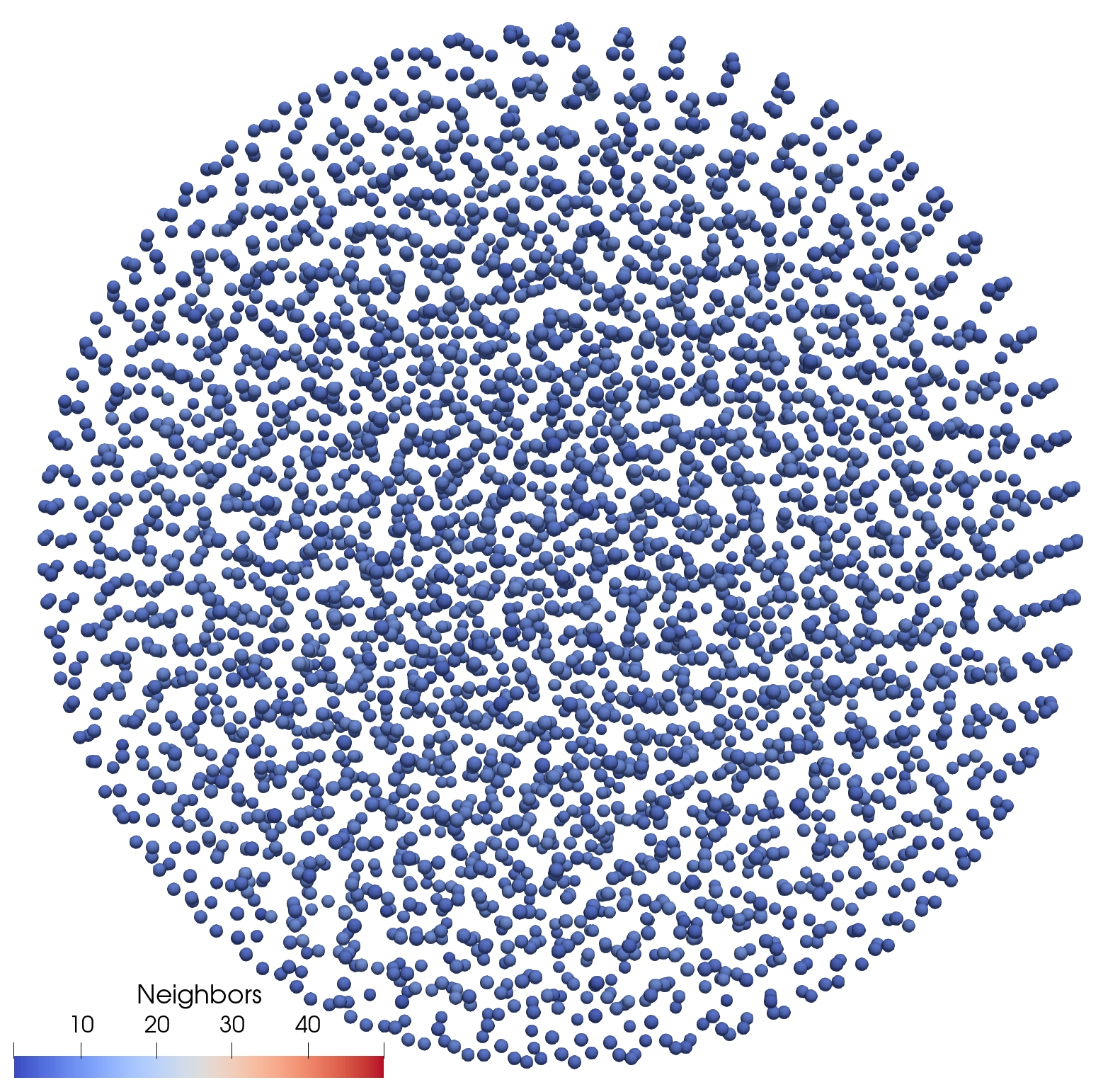}}
    \subfigure[$n$=8,000]
    {\includegraphics[width = 0.45\textwidth]{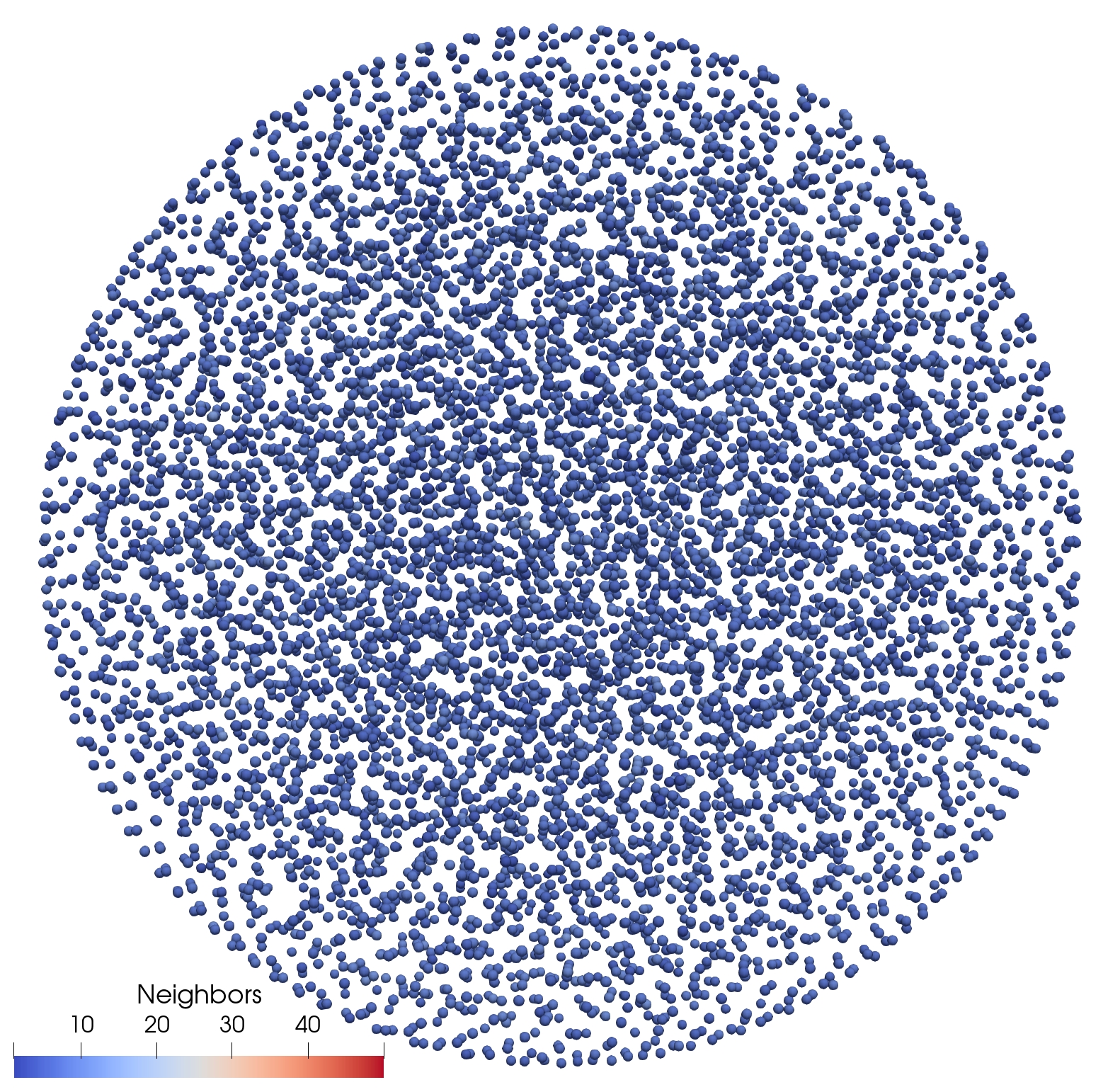}}
    \subfigure[$n$=16,000]
    {\includegraphics[width = 0.45\textwidth]{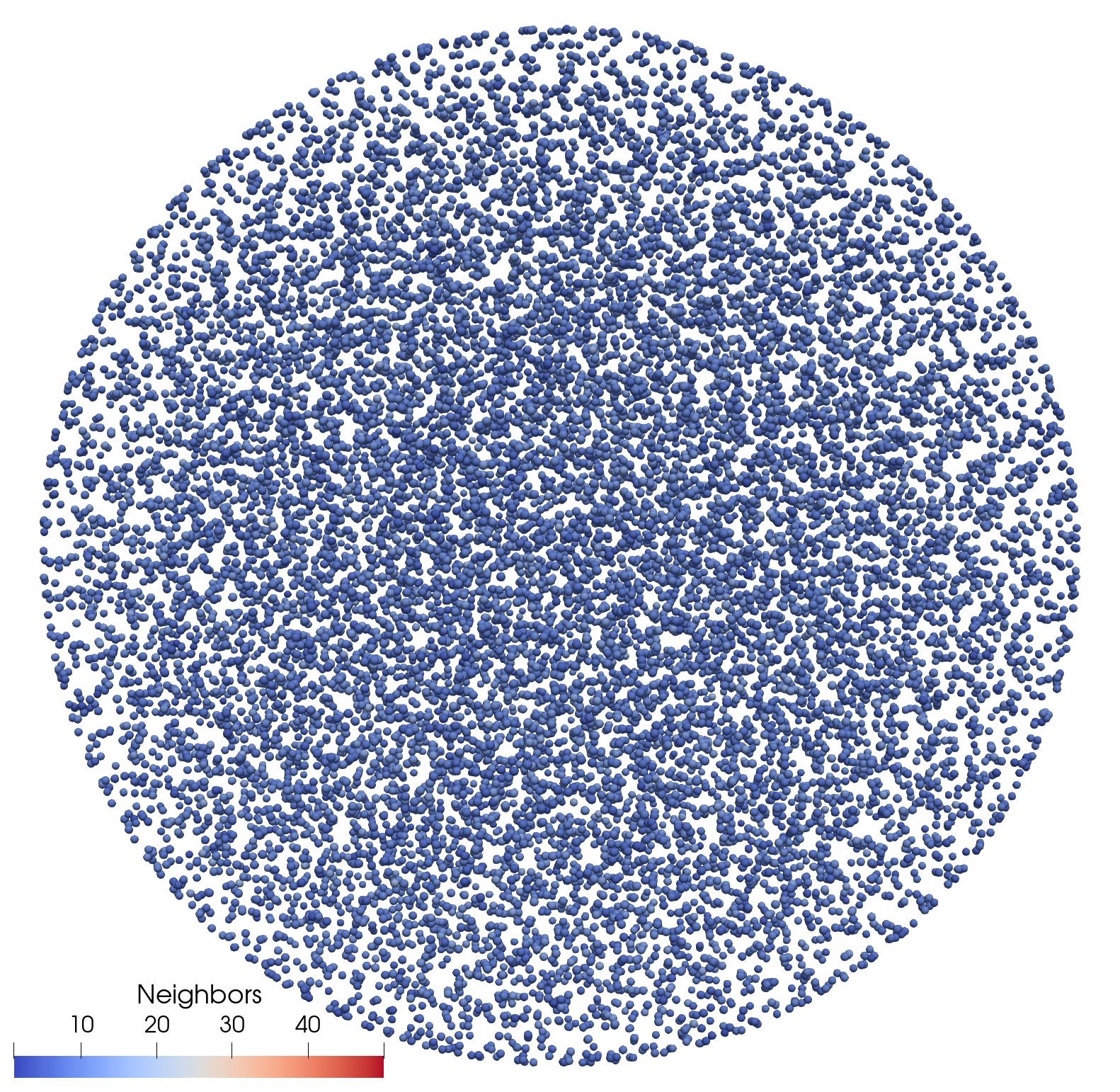}}
    \subfigure[$n$=32,000]
    {\includegraphics[width = 0.45\textwidth]{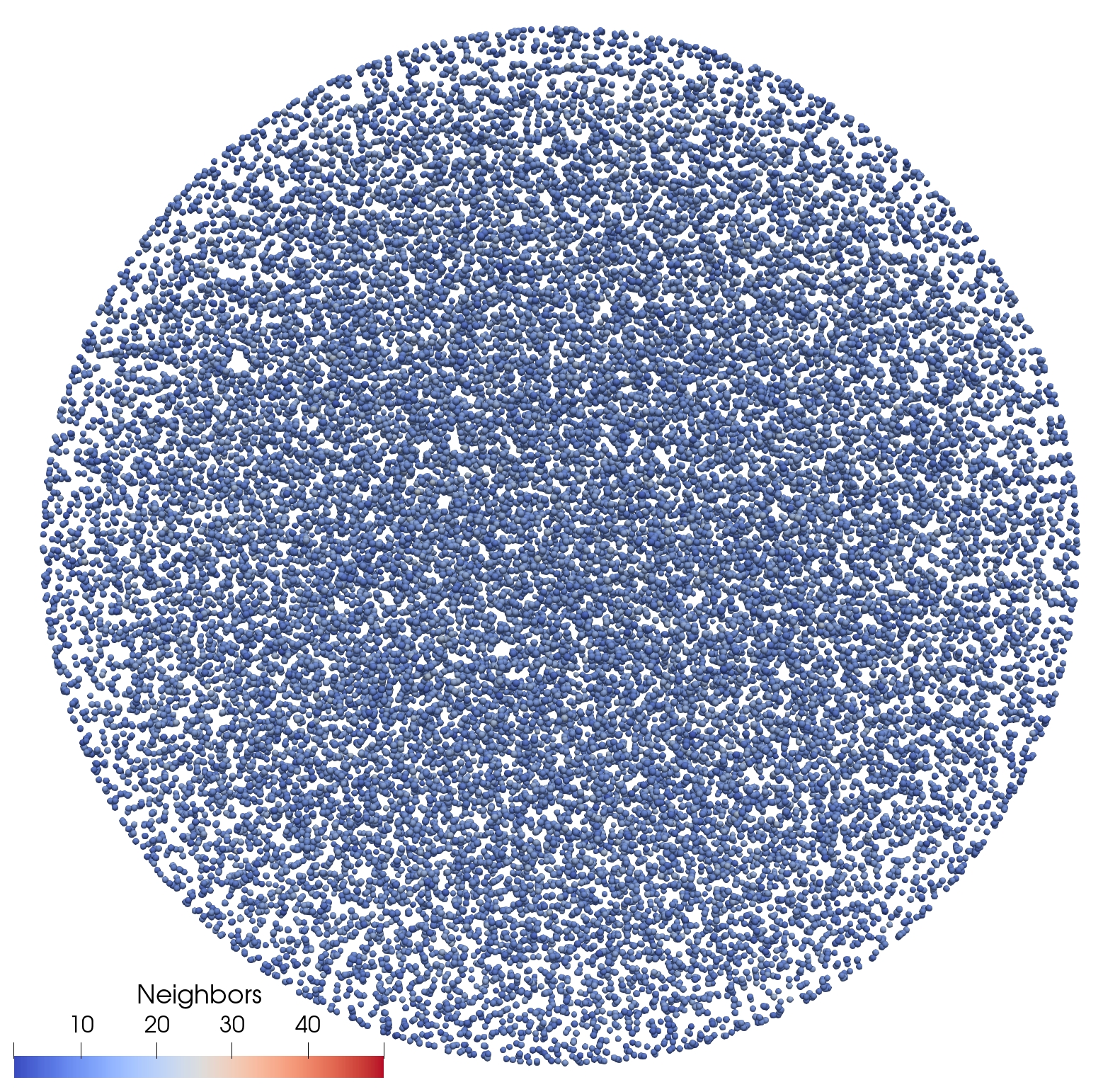}}
    \caption{\footnotesize Visualization of the 3D distribution of the centers of the particles at the end of the simulation time $T$ for different number of particles. The analysis adopts a constant value of $\beta$ pegged through the optimal scaling relation to the number of particles $n$. The color of each particle relates to the number of particles in its neighborhood. The size of the particles is related to their mass $m=M/n$.}
    \label{Fig:analysisEnd}
\end{figure}

\begin{figure}[!h]
    \centering
    \subfigure[$n$=4,000]
    {\includegraphics[width = 0.45\textwidth]{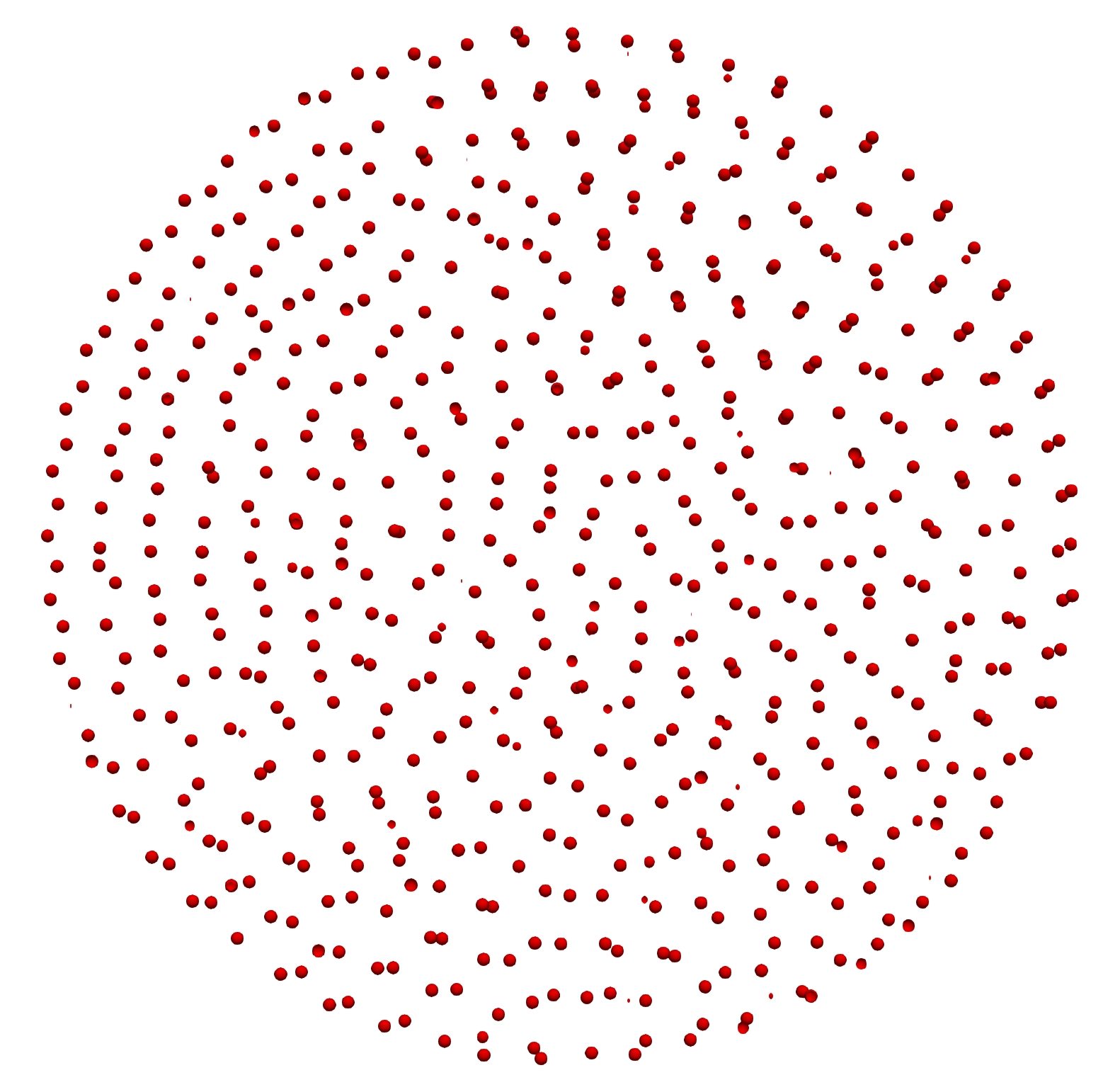}}
    \subfigure[$n$=8,000]
    {\includegraphics[width = 0.45\textwidth]{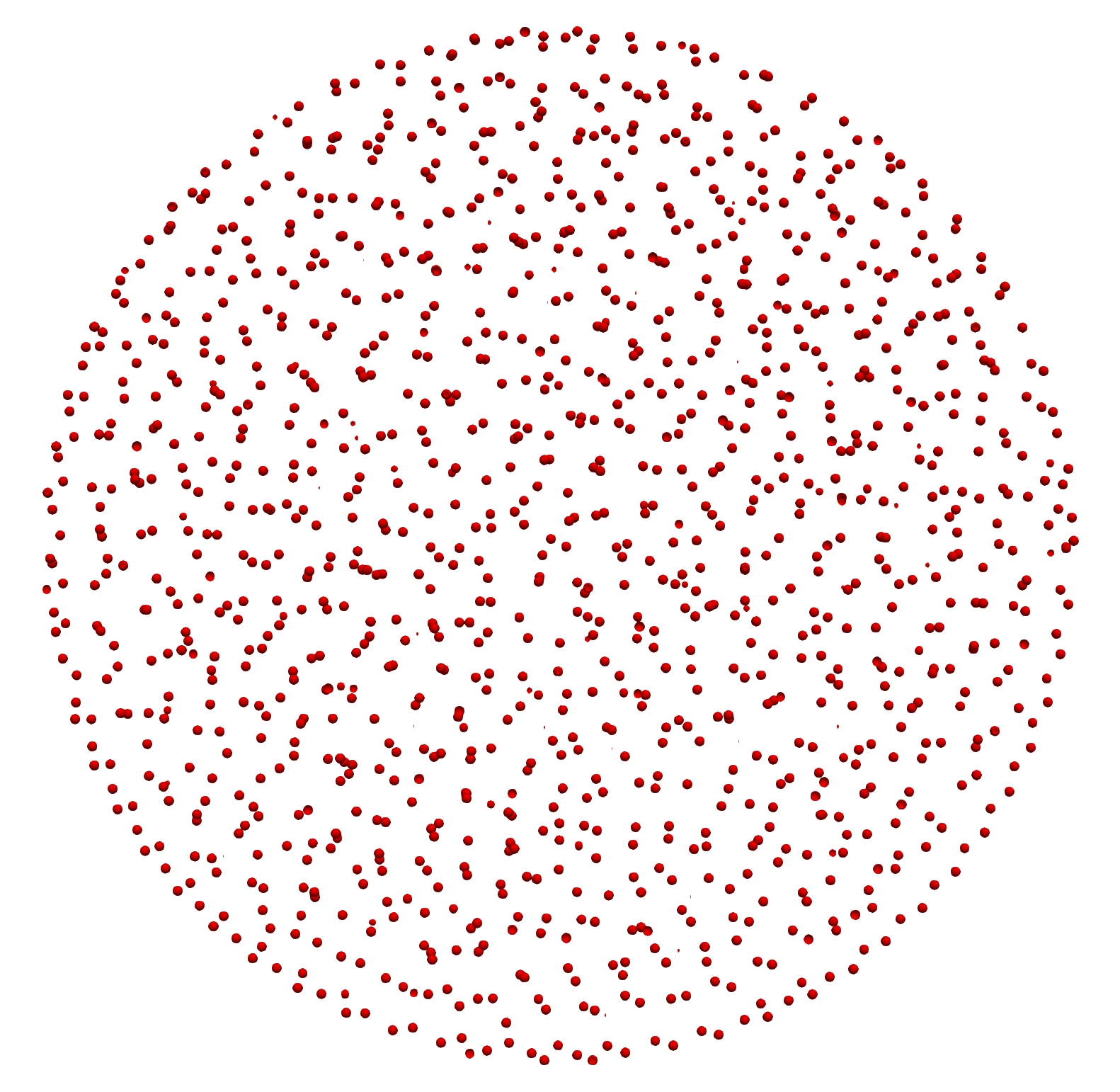}}
    \subfigure[$n$=16,000]
    {\includegraphics[width = 0.45\textwidth]{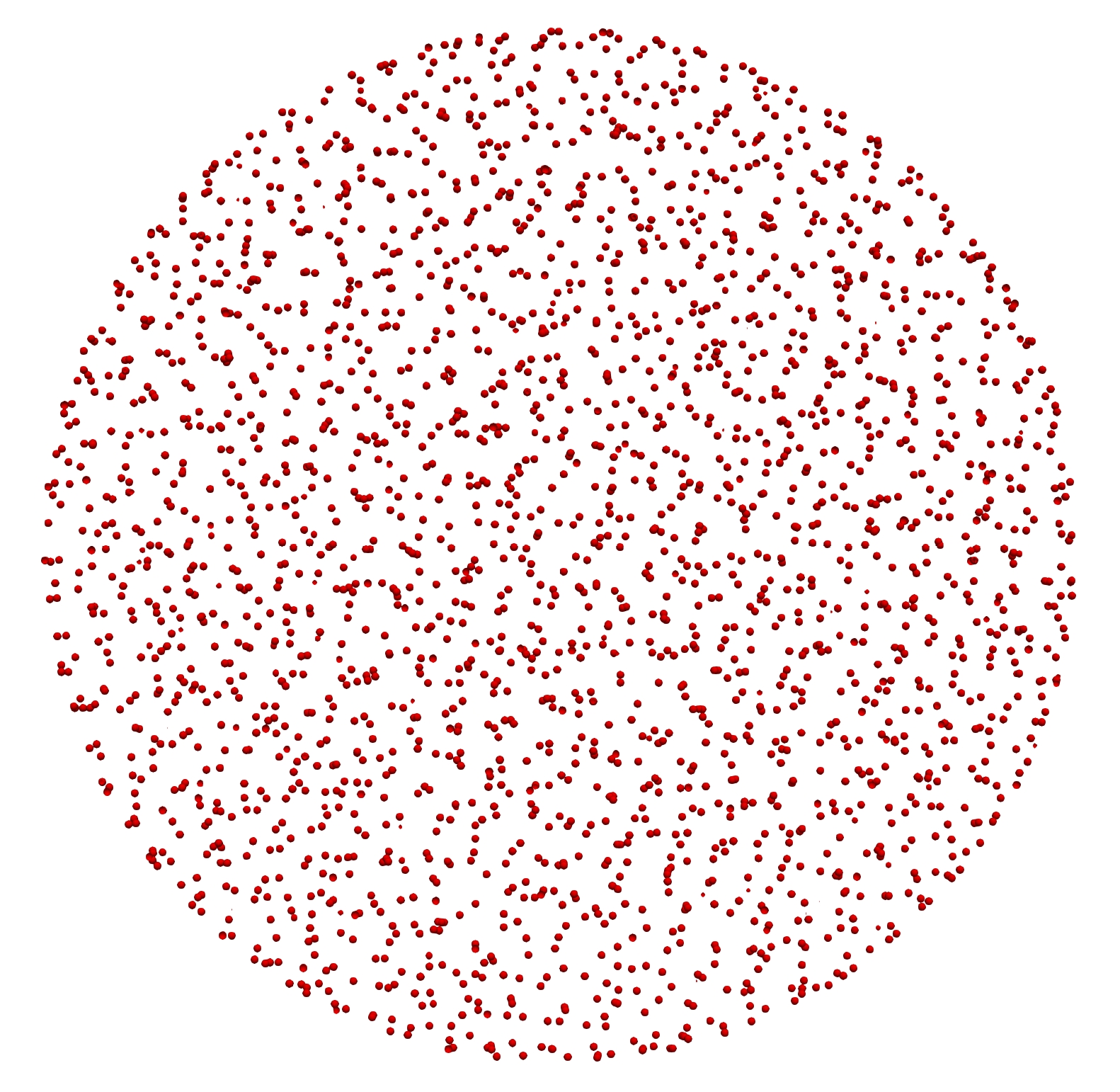}}
    \subfigure[$n$=32,000]
    {\includegraphics[width = 0.45\textwidth]{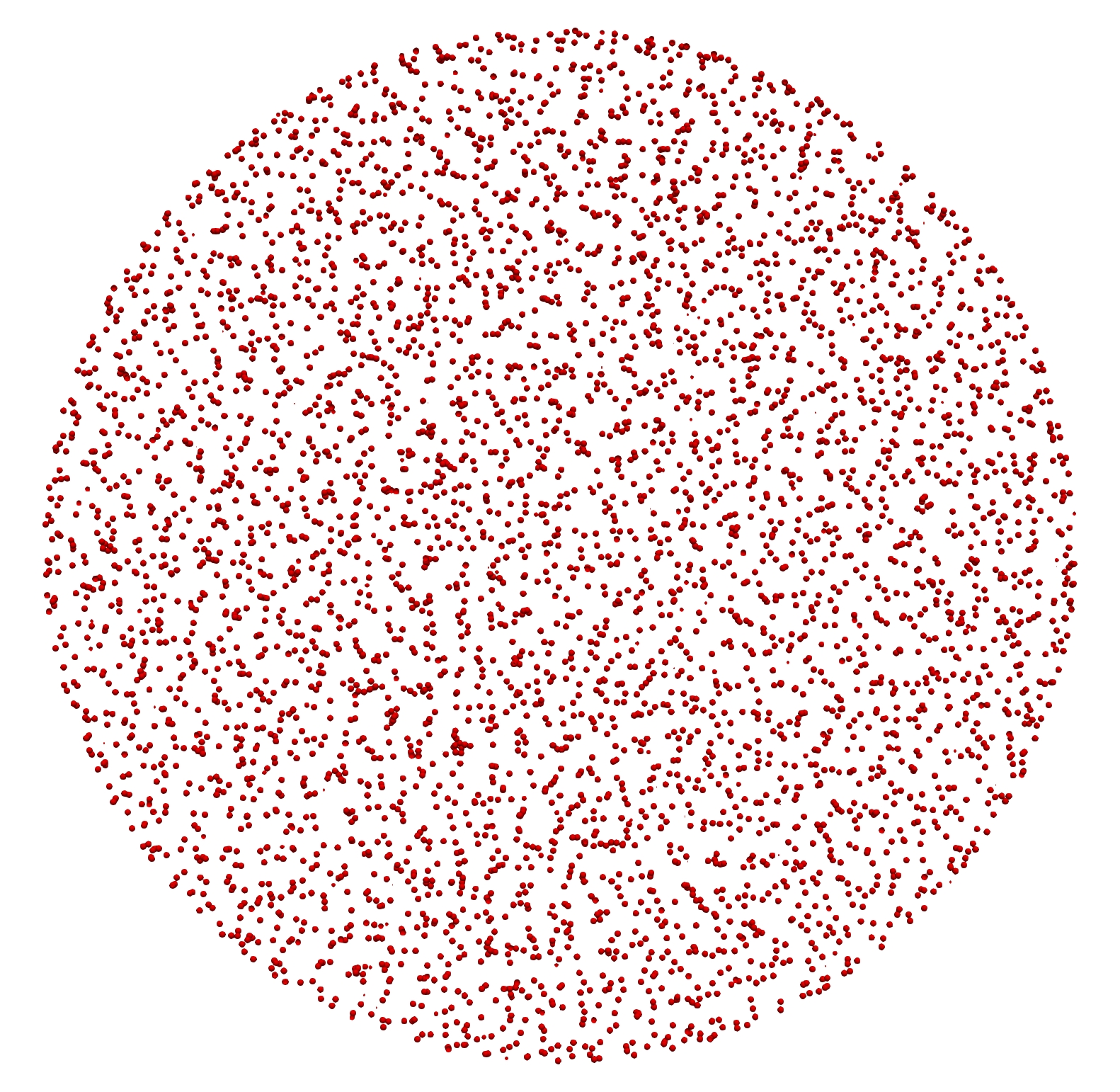}}
    \caption{\footnotesize Visualization of the distribution of the particles at the end of the analysis for different number of particles, $C_3=10$. The image shows all the particles contained within a meridian disk-like slice of the sphere, 0.2 thick, taken at the maximum diameter. The particles have the same size, but the associated mass is different in the four cases.}
    \label{Fig:maximumCircleDistribution}
\end{figure}

Fig.~\ref{Fig:analysisEnd} shows the same distributions of the centers of the particles at the end of the simulation time $T$ for different values of $n$. We recall that the slowest relaxation time of the diffusional problem scales as $R^2/\kappa = 1$. At $T=120$, we may therefore expect the system to be at steady state. Indeed, an ostensibly uniform steady-state density is clear from the figures, as expected. The uniformity of the steady-state density is borne out by a different view of the system, Fig.~\ref{Fig:maximumCircleDistribution}, which depicts a meridional section of the sphere in the form of a disk of width 0.2. The long-term stability of the finite-particle scheme is again remarkable. 

\begin{figure}
    \centering
    \subfigure[$n$=1000]
    {\includegraphics[width = 0.7\textwidth]{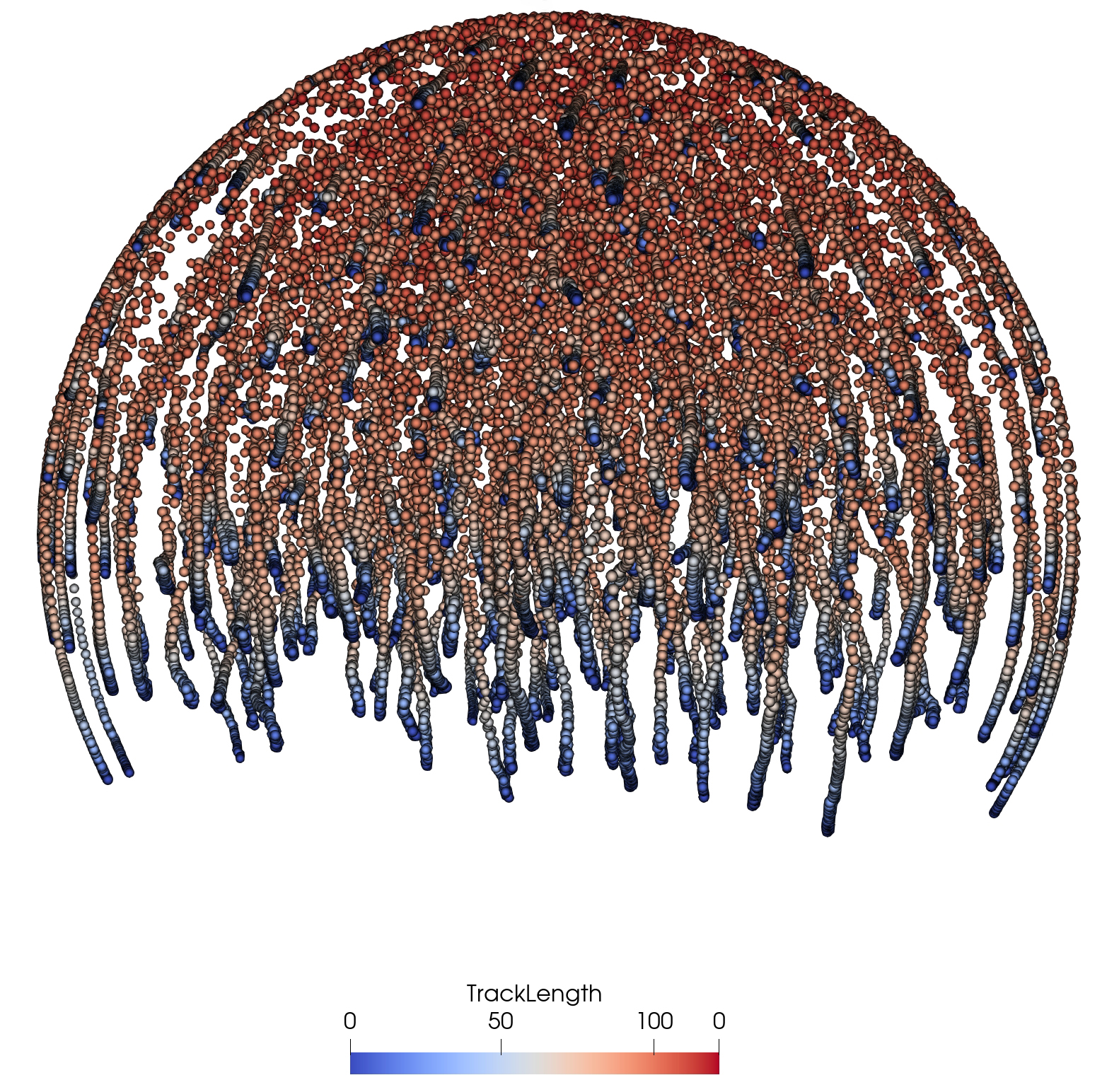}}
    \subfigure[$n$=4000]
    {\includegraphics[width = 0.7\textwidth]{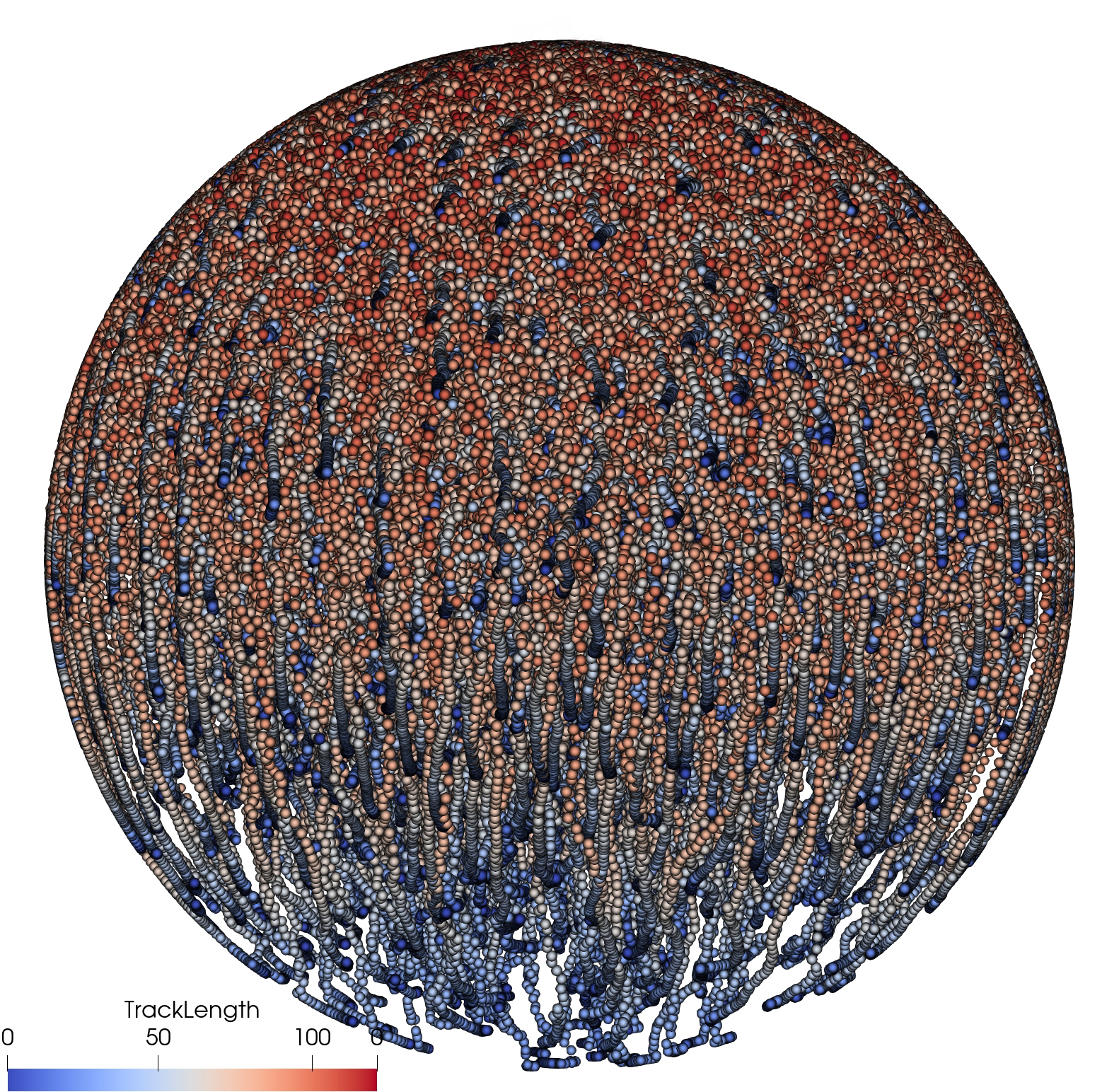}}
    \caption{\footnotesize Long-exposure picture of the flow showing the particle streamers, or paths. The particles fill the domain following injection and trace trajectories tangential to the spherical surface of the domain, in accordance with the zero-flux boundary condition.}
    \label{Fig:PathLines}
\end{figure}

A different type of visualization, which supplies a sense of the motion of the particles, consists of simulating a long-exposure picture of the flow showing the particle streamers, or paths. In Fig.~\ref{Fig:PathLines}, these streamers are shown for $n = 1,000$ and $4,000$ only, since larger numbers of particles result in excessive clutter. It is noteworthy from the figures how the particles spread out from the injection point and gradually fill the sphere. In particular, the outermost particles trace trajectories tangential to the spherical surface of the domain, as enforced by the barrier potential, thus abiding by the zero-flux boundary condition. 

\subsection{Analysis of convergence}

We assess the (weak) convergence of the method by computing the polar moment of inertia of the distribution at steady-state, 
\begin{equation} \label{eq:analyticalInertia}
    J_G = \frac{3}{5} M R^2 \, ,
    \qquad 
    M = \frac{4}{3} \pi R^3 \rho_\infty \, . 
\end{equation}
The approximate value from the particle discretization is 
\begin{equation}
    \label{eq:numericalInertia}
    J_G = \sum_{p=1}^n J_p \, ,
\end{equation}
where
\begin{equation} \label{eq:singleParticleAxis}
    J_p 
    =  
    m_p 
    \int_\Omega 
        r^2 {N}(x - x_p; \beta ) 
    \, dx 
    =
    m_p \Big( r_p^2 + \frac{2}{3 \beta } \Big) 
\end{equation}
and we write $r_p=\|x_p\|$.

\begin{figure}[!h]
    \centering
    \centering
    \subfigure[]
    {\includegraphics[width = 0.7\textwidth]{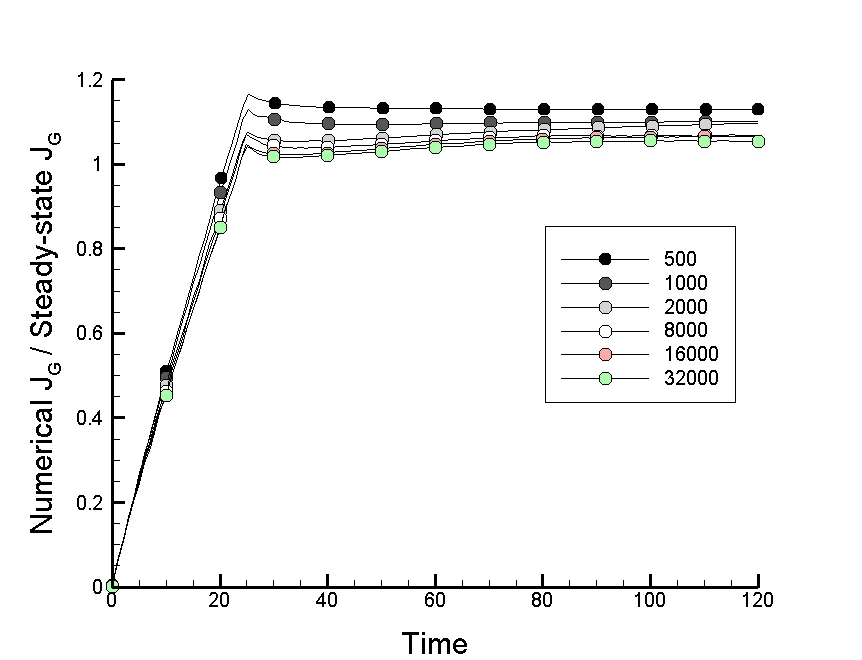}}
    \subfigure[]
    {\includegraphics[width = 0.7\textwidth]{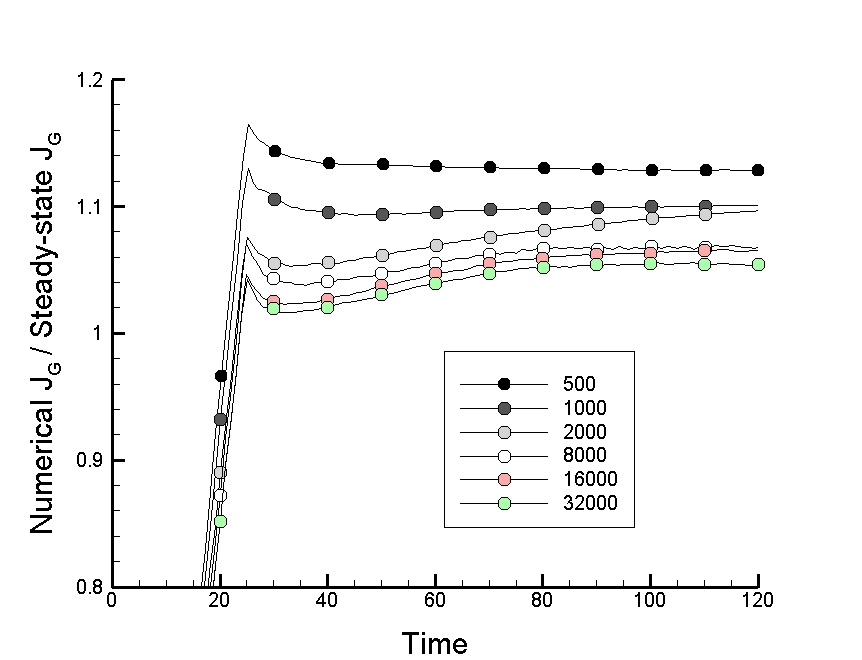}}
    \caption{\footnotesize Time evolution of the numerical value of the polar momentum of inertia, normalized with respect to the steady-state value, for various numbers of particles. (a) Full history. (b) Detail of the attainment of the steady state condition.}
    \label{Fig:JGevolution}
\end{figure}

The time evolution of the numerical polar moment of inertia normalized with respect to the steady-state value \eqref{eq:analyticalInertia} is shown in Fig.~\ref{Fig:JGevolution} as a function of the number of particles. The initial injection phase and the subsequent diffusion phase are clearly evident in the figure. The polar moment of inertia overshoots the steady state value at the end of the injection phase and then converges to the steady state from above in at large times, as expected. A general trend towards convergence of the time evolution---and towards the steady-state a large times---with respect to the number of particles is also clearly visible in the figure.

\begin{figure}[!h]
    \centering
    \subfigure[Runge-Kutta]
    {\includegraphics[width = 0.7\textwidth]{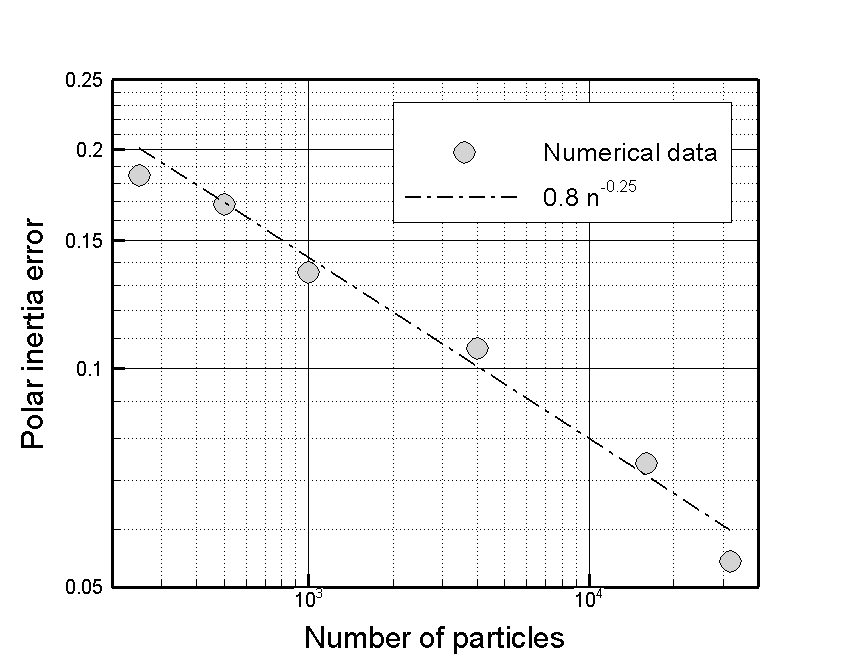}}
    \subfigure[Forward-Euler]
    {\includegraphics[width = 0.7\textwidth]{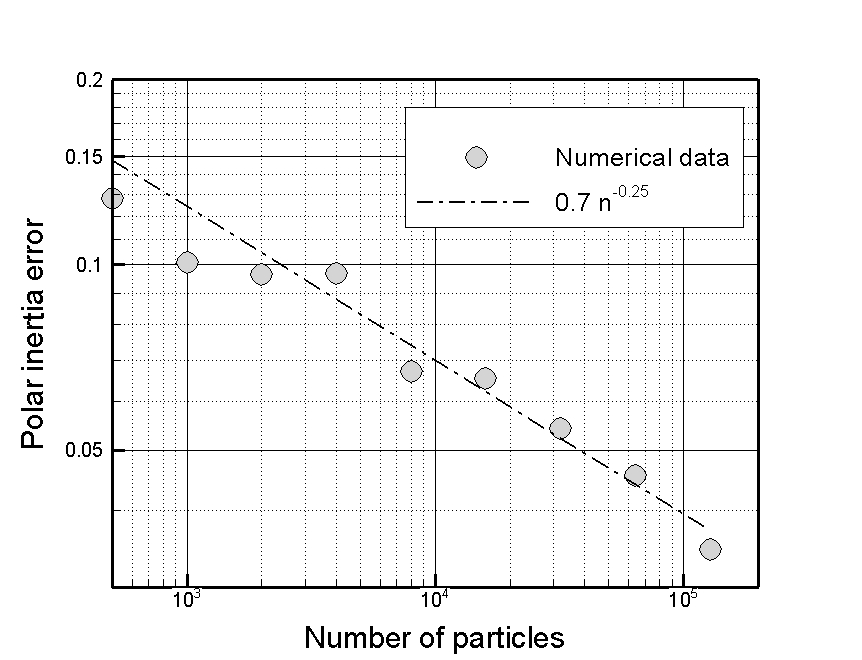}}
    \caption{\footnotesize Relative error of the polar inertia as a function of the number of particles and linear regression. a) Results using Runge-Kutta integration, sequential. b) Results using the Forward Euler integration, parallel, with increasing number of cores. The inferred rate of convergence is $1/4$ for both cases.}
    \label{Fig:InertiaMoment}
\end{figure}

The relative error of the numerical polar moment of inertia at steady-state is plotted in Fig.~\ref{Fig:InertiaMoment} as a function of the number of particles. This error effectively quantifies the spatial accuracy of the finite-particle discretization, including the clipping of the particle tail profiles at the boundary, for a random Poisson distribution of particle centers. As expected (cf., e.~g., \cite{Huesmann:2013, Schmidt:2014}), the plots show a clear trend towards convergence at a rate of convergence of $1/4$. 

\section{Summary and concluding remarks}
\label{sec:conclusions}

We have formulated a class of finite-particle methods for mass transport problems based on a time-discrete incremental variational principle that combines entropy and the cost of particle transport, as measured by the Wasserstein metric. The incremental functional is further spatially discretized into finite particles, i.~e., particles characterized by a fixed spatial profile of finite width, each carrying a fixed amount of mass. The motion of the particles is then governed by a competition between the cost of transport, that aims to keep the particles fixed, and entropy maximization, that aims to spread the particles so as to increase the entropy of the system. Unlike other particle schemes based on optimal transport and intended for deformable bodies \cite{LiHabbalOrtiz2010, fedeli2017geometrically, Navas:2018, Wessels:2018}, the formulation is velocity-free, i.~e., it relies solely on the particle representation and eschews the need to define a velocity field in addition, which confers the approach great advantage. 

A critical issue in formulating finite-particle methods concerns the optimal choice of particle width. We have shown how the optimal width of the particles can be determined variationally by minimization of the governing incremental functional. In this manner, the incremental functional determines not only the motion of the particles but also the evolution of their profiles. Using this extended variational principle, we have in particular derived optimal scaling relations between the width of the particles, their number and the size of the domain. It is found that the optimal particle width lies in an intermediate scale between the particle distance and the size of the domain, with exponents that are precisely determined by the theory. 

We have also addressed sundry matters of implementation including the acceleration of the computation of diffusive forces by exploiting the Gaussian decay of the particle profiles and by instituting fast nearest-neighbor searches. We have demonstrated how a simple domain subdivision scheme, with particle-neighborhood size and cell sizes subordinate to the optimal particle profile width, greatly accelerates the computation of diffusional forces, which is the main computational bottleneck, and allows consideration of large systems of particles. 

Finally, we have demonstrated the robustness and versatility of the finite-particle method by means of a test problem concerned with the injection of mass into a sphere. There test results demonstrate the meshless character of the method in any spatial dimension, its ability to redistribute mass particles and follow their evolution in time, its ability to satisfy flux boundary conditions for general domains based solely on a distance function, and its robust convergence characteristics.

We close by remarking that the finite-particle method presented in this work, and others like it, is not restricted to mass transport and can in principle be applied to other scalar transport problems. However, whereas Dirichlet boundary conditions are somewhat inimical to mass transport, they play a central role in other transport problems such as heat conduction. In the present work, we have eschewed consideration of Dirichlet boundary conditions in the interest of conciseness, but it remains of interest to elucidate efficient implementations thereof. In addition, while the simple one-level subdivision scheme employed in this work has sufficed to explore the convergence properties of the method, it remains of interest to explore other fast-search algorithms, including hierarchical $k$-means, approximate nearest-neighbor (ANN) algorithms and graph-based methods. These and other extensions and enhancements of the approach suggest themselves as worthwhile directions for further research. 

\section*{Acknowledgements}
AP is grateful for support of the Italian National Group of Physics-Mathematics (GNFM) of the Italian National Institution of High Mathematics ‘‘Francesco Severi’’ (INDAM). MO gratefully acknowledges support from the Deutsche Forschungsgemeinschaft (DFG) through the Schwerpunkt-\linebreak programme (SPP) 2311. 

\bibliography{ms}
\bibliographystyle{unsrt}

\end{document}